\documentclass[twocolumn]{aastex631}
\usepackage{hyperref}
\usepackage{float}
\usepackage{threeparttable} 
\usepackage{captdef} 
\usepackage{tabularx} 

\newcommand{\brad}{2.99}
\newcommand{\braderr}{0.14}
\newcommand{\bmass}{10.8} 
\newcommand{\bmasserr}{1.8}
\newcommand{\bTSM}{108.8}
\newcommand{\bTSMerr}{35.8}
\newcommand{\bESM}{10.5}
\newcommand{\bESMerr}{1.6}

\newcommand{\bdensity}{2.2}
\newcommand{\bdensityerr}{0.5}

\newcommand{\crad}{2.90}
\newcommand{\craderr}{0.13}
\newcommand{\cmass}{11.1} 
\newcommand{\cmasserr}{2.4}
\newcommand{\cTSM}{72.5}
\newcommand{\cTSMerr}{21.3}
\newcommand{\cESM}{4.8}
\newcommand{\cESMerr}{0.7}

\newcommand{\cdensity}{2.5}
\newcommand{\cdensityerr}{0.6} 

\newcommand{\mearth}{$M_\oplus$}
\newcommand{\rearth}{$R_\oplus$}

\newcommand{\tess}{TESS }
\newcommand{\jwst}{\textit{JWST }}
\newcommand{\gaia}{\textit{Gaia }}

\shorttitle{TKS V. Twin sub-Neptunes Transiting the Nearby G Star HD 63935}
\shortauthors{Scarsdale et. al.}

\begin{document}


\title{TESS-Keck Survey V. Twin sub-Neptunes Transiting the Nearby G Star HD 63935}


\author[0000-0003-3623-7280]{Nicholas Scarsdale}
\affiliation{Department of Astronomy and Astrophysics, University of California, Santa Cruz, CA 95064, USA}

\author[0000-0001-8898-8284]{Joseph M. Akana Murphy} 
\altaffiliation{NSF Graduate Research Fellow}
\affiliation{Department of Astronomy and Astrophysics, University of California, Santa Cruz, CA 95064, USA}

\author[0000-0002-7030-9519]{Natalie M. Batalha}
\affiliation{Department of Astronomy and Astrophysics, University of California, Santa Cruz, CA 95064, USA}


\author{Ian J. M. Crossfield}
\affiliation{Department of Physics \& Astronomy, University of Kansas, 1082 Malott,1251 Wescoe Hall Dr., Lawrence, KS 66045, USA}

\author[0000-0001-8189-0233]{Courtney D. Dressing}
\affiliation{Department of Astronomy,  University of California Berkeley, Berkeley CA 94720, USA}

\author[0000-0003-3504-5316]{Benjamin Fulton}
\affiliation{NASA Exoplanet Science Institute/Caltech-IPAC, MC 314-6, 1200 E. California Blvd., Pasadena, CA 91125, USA}

\author[0000-0001-8638-0320]{Andrew W. Howard}
\affiliation{Department of Astronomy, California Institute of Technology, Pasadena, CA 91125, USA}

\author[0000-0001-8832-4488]{Daniel Huber}
\affiliation{Institute for Astronomy, University of Hawai`i, 2680 Woodlawn Drive, Honolulu, HI 96822, USA}

\author[0000-0002-0531-1073]{Howard Isaacson}
\affiliation{Department of Astronomy,  University of California Berkeley, Berkeley CA 94720, USA}
\affiliation{Centre for Astrophysics, University of Southern Queensland, Toowoomba, QLD, Australia}

\author[0000-0002-7084-0529]{Stephen R. Kane}
\affiliation{Department of Earth and Planetary Sciences, University of California, Riverside, CA 92521, USA}

\author[0000-0003-0967-2893]{Erik A Petigura}
\affiliation{Department of Physics \& Astronomy, University of California Los Angeles, Los Angeles, CA 90095, USA}

\author[0000-0003-0149-9678]{Paul Robertson}
\affiliation{Department of Physics \& Astronomy, University of California Irvine, Irvine, CA 92697, USA}

\author[0000-0001-8127-5775]{Arpita Roy}
\affiliation{Space Telescope Science Institute, 3700 San Martin Drive, Baltimore, MD 21218, USA}
\affiliation{Department of Physics and Astronomy, Johns Hopkins University, 3400 N Charles St, Baltimore, MD 21218, USA}

\author[0000-0002-3725-3058]{Lauren M. Weiss}
\affiliation{Institute for Astronomy, University of Hawai`i, 2680 Woodlawn Drive, Honolulu, HI 96822, USA}


\author[0000-0001-7708-2364]{Corey Beard}
\affiliation{Department of Physics \& Astronomy, University of California Irvine, Irvine, CA 92697, USA}

\author[0000-0003-0012-9093]{Aida Behmard}
\altaffiliation{NSF Graduate Research Fellow}
\affiliation{Division of Geological and Planetary Science, California Institute of Technology, Pasadena, CA 91125, USA}

\author[0000-0003-1125-2564]{Ashley Chontos}
\altaffiliation{NSF Graduate Research Fellow}
\affiliation{Institute for Astronomy, University of Hawai`i, 2680 Woodlawn Drive, Honolulu, HI 96822, USA}

\author[0000-0002-8035-4778]{Jessie L. Christiansen}
\affiliation{NASA Exoplanet Science Institute-Caltech/IPAC, 1200 E. California Blvd, Pasadena, CA 91125 USA}

\author[0000-0002-5741-3047]{David R. Ciardi}
\affiliation{NASA Exoplanet Science Institute-Caltech/IPAC, 1200 E. California Blvd, Pasadena, CA 91125 USA}

\author[0000-0002-9879-3904]{Zachary R. Claytor}
\affiliation{Institute for Astronomy, University of Hawai`i, 2680 Woodlawn Drive, Honolulu, HI 96822, USA}

\author[0000-0001-6588-9574]{Karen A.\ Collins}
\affiliation{Center for Astrophysics | Harvard \& Smithsonian, 60 Garden Street, Cambridge, Massachusetts 02138, USA}

\author[0000-0003-2781-3207]{Kevin I.\ Collins}
\affiliation{George Mason University, 4400 University Drive, Fairfax, VA, 22030 USA}

\author[0000-0002-8958-0683]{Fei Dai}
\affiliation{Division of Geological and Planetary Science, California Institute of Technology, Pasadena, CA 91125, USA}

\author[0000-0002-4297-5506]{Paul A. Dalba}
\altaffiliation{NSF Astronomy and Astrophysics Postdoctoral Fellow}
\affiliation{Department of Earth and Planetary Sciences, University of California, Riverside, CA 92521, USA}

\author[0000-0003-2313-467X]{Diana~Dragomir}
\affiliation{Department of Physics and Astronomy, University of New Mexico, 210 Yale Blvd NE, Albuquerque, NM 87106, USA}

\author[0000-0002-3551-279X]{Tara Fetherolf}
\affiliation{Department of Earth and Planetary Sciences, University of California, Riverside, CA 92521, USA}

\author[0000-0002-4909-5763]{Akihiko Fukui}
\affiliation{Komaba Institute for Science, The University of Tokyo, 3-8-1 Komaba, Meguro, Tokyo 153-8902, Japan}
\affiliation{Instituto de Astrof\'{i}sica de Canarias (IAC), 38205 La Laguna, Tenerife, Spain}

\author[0000-0002-8965-3969]{Steven Giacalone}
\affiliation{Department of Astronomy,  University of California Berkeley, Berkeley CA 94720, USA}

\author{Erica J. Gonzales}
\altaffiliation{NSF Graduate Research Fellow}
\affiliation{Department of Astronomy and Astrophysics, University of California, Santa Cruz, CA 95064, USA}

\author[0000-0002-0139-4756]{Michelle L. Hill}
\affiliation{Department of Earth and Planetary Sciences, University of California, Riverside, CA 92521, USA}

\author[0000-0001-8058-7443]{Lea A.\ Hirsch}
\affiliation{Kavli Institute for Particle Astrophysics and Cosmology, Stanford University, Stanford, CA 94305, USA}

\author[0000-0002-4625-7333]{Eric L.\ N.\ Jensen}
\affiliation{Department of Physics \& Astronomy, Swarthmore College, Swarthmore PA 19081, USA}

\author[0000-0002-6115-4359]{Molly R. Kosiarek}
\altaffiliation{NSF Graduate Research Fellow} 
\affiliation{Department of Astronomy and Astrophysics, University of California, Santa Cruz, CA 95064, USA}

\author[0000-0002-6424-3410]{Jerome P. de Leon}
\affiliation{Department of Astronomy, Graduate School of Science, The University of Tokyo, 7-3-1 Hongo, Bunkyo-ku, Tokyo 113-0033, Japan}

\author[0000-0001-8342-7736]{Jack Lubin}
\affiliation{Department of Physics \& Astronomy, University of California Irvine, Irvine, CA 92697, USA}

\author[0000-0003-2527-1598]{Michael B. Lund}
\affiliation{NASA Exoplanet Science Institute-Caltech/IPAC, 1200 E. California Blvd, Pasadena, CA 91125 USA}

\author[0000-0002-4671-2957]{Rafael Luque}
\affiliation{Instituto de Astrof\'isica de Andaluc\'ia (IAA-CSIC), Glorieta de la Astronom\'ia s/n, 18008 Granada, Spain}

\author{Andrew W. Mayo}
\affiliation{Department of Astronomy,  University of California Berkeley, Berkeley CA 94720, USA}
\affiliation{Centre for Star and Planet Formation, Natural History Museum of Denmark \& Niels Bohr Institute, University of Copenhagen, Øster Voldgade 5-7, DK-1350 Copenhagen K., Denmark}

\author[0000-0003-4603-556X]{Teo Mo\v{c}nik}
\affiliation{Gemini Observatory/NSF's NOIRLab, 670 N. A'ohoku Place, Hilo, HI 96720, USA}

\author[0000-0003-1368-6593]{Mayuko Mori}
\affiliation{Department of Astronomy, Graduate School of Science, The University of Tokyo, 7-3-1 Hongo, Bunkyo-ku, Tokyo 113-0033, Japan}

\author[0000-0001-8511-2981]{Norio Narita}
\affiliation{Komaba Institute for Science, The University of Tokyo, 3-8-1 Komaba, Meguro, Tokyo 153-8902, Japan}
\affiliation{Japan Science and Technology Agency, PRESTO, 3-8-1 Komaba, Meguro, Tokyo 153-8902, Japan}
\affiliation{Astrobiology Center, 2-21-1 Osawa, Mitaka, Tokyo 181-8588, Japan}
\affiliation{Instituto de Astrof\'{i}sica de Canarias (IAC), 38205 La Laguna, Tenerife, Spain}

\author[0000-0002-7031-7754]{Grzegorz Nowak}
\affiliation{Instituto de Astrof\'\i sica de Canarias (IAC), 38205 La Laguna, Tenerife, Spain}
\affiliation{Departamento de Astrof\'\i sica, Universidad de La Laguna (ULL), 38206 La Laguna, Tenerife, Spain}

\author[0000-0003-0987-1593]{Enric~Pall\'e}
\affiliation{Instituto de Astrof\'\i sica de Canarias (IAC), 38205 La Laguna, Tenerife, Spain}
\affiliation{Departamento de Astrof\'\i sica, Universidad de La Laguna (ULL), 38206 La Laguna, Tenerife, Spain}

\author{Markus~Rabus}
\affiliation{Departamento de Matem\'atica y F\'isica Aplicadas, Universidad Cat\'olica de la Sant\'isima Concepci\'on, Alonso de Rivera 2850, Concepci\'on, Chile}
\affiliation{Las Cumbres Observatory Global Telescope Network, Santa Barbara, CA 93117, USA}
\affiliation{Department of Physics, University of California, Santa Barbara, CA 93106-9530, USA}

\author{Lee J.\ Rosenthal}
\affiliation{Department of Astronomy, California Institute of Technology, Pasadena, CA 91125, USA}

\author[0000-0003-3856-3143]{Ryan A. Rubenzahl}
\altaffiliation{NSF Graduate Research Fellow}
\affiliation{Department of Astronomy, California Institute of Technology, Pasadena, CA 91125, USA}

\author{Joshua E. Schlieder}
\affiliation{NASA Goddard Space Flight Center, Greenbelt, MD 20771, USA}

\author[0000-0002-1836-3120]{Avi~Shporer}
\affiliation{Department of Physics and Kavli Institute for Astrophysics and Space Research, Massachusetts Institute of Technology, Cambridge, MA 02139, USA}

\author[0000-0002-3481-9052]{Keivan G. Stassun}
\affiliation{Department of Physics \& Astronomy, Vanderbilt University, Nashville, TN, USA}

\author[0000-0002-6778-7552]{Joe Twicken}
\affiliation{NASA Ames Research Center, Moffett Field, CA 94035, USA}
\affiliation{SETI Institute, Mountain View, CA 94043, USA}

\author[0000-0003-3092-4418]{Gavin Wang}
\affiliation{Tsinghua International School, Beijing 100084, China}

\author[0000-0002-5402-9613]{Bill Wohler}
\affiliation{NASA Ames Research Center, Moffett Field, CA 94035, USA}
\affiliation{SETI Institute, Mountain View, CA 94043, USA}

\author[0000-0003-4755-584X]{Daniel A. Yahalomi}
\affiliation{Department of Astronomy, Columbia University, 550 W 120th St., New York NY 10027, USA}

\author[0000-0002-4715-9460]{Jon Jenkins} 
\affil{NASA Ames Research Center, Moffett Field, CA 94035, USA}

\author[0000-0001-9911-7388]{David W. Latham} 
\affil{Center for Astrophysics | Harvard \& Smithsonian, 60 Garden Street, Cambridge, Massachusetts 02138, USA}

\author[0000-0003-2058-6662]{George R. Ricker}
\affil{Department of Physics and Kavli Institute for Astrophysics and Space Research, Massachusetts Institute of Technology, Cambridge, MA 02139, USA}

\author[0000-0002-6892-6948]{S.~Seager}
\affiliation{Department of Physics and Kavli Institute for Astrophysics and Space Research, Massachusetts Institute of Technology, Cambridge, MA 02139, USA}
\affiliation{Department of Earth, Atmospheric and Planetary Sciences, Massachusetts Institute of Technology, Cambridge, MA 02139, USA}
\affiliation{Department of Aeronautics and Astronautics, MIT, 77 Massachusetts Avenue, Cambridge, MA 02139, USA}

\author{Roland Vanderspek} 
\affil{Department of Physics and Kavli Institute for Astrophysics and Space Research, Massachusetts Institute of Technology, Cambridge, MA 02139, USA}

\author[0000-0002-4265-047X]{Joshua N.\ Winn}
\affiliation{Department of Astrophysical Sciences, Peyton Hall, 4 Ivy Lane, Princeton, NJ 08544, USA}


\begin{abstract} 

We present the discovery of two nearly identically-sized sub-Neptune transiting planets orbiting HD 63935, a bright ($V=8.6$ mag), sun-like ($T_{eff}=5560K$) star at 49 pc. 
\tess identified the first planet, HD 63935 b (TOI-509.01), in Sectors 7 and 34.
We identified the second signal (HD 63935 c) in Keck HIRES and Lick APF radial velocity data as part of our followup campaign. It was subsequently confirmed with \tess photometry in Sector 34 as TOI-509.02.
Our analysis of the photometric and radial velocity data yields a robust detection of both planets with periods of $9.0600 \pm 0.007$ and $21.40 \pm 0.0019$ days, radii of $\brad \pm \braderr$ and $\crad \pm \craderr$ $R_\oplus$, and masses of $\bmass \pm \bmasserr$ and $\cmass \pm \cmasserr$ \mearth.
We calculate densities for planets b and c consistent with a few percent of the planet mass in hydrogen/helium envelopes.
We also describe our survey's efforts to choose the best targets for \jwst atmospheric followup.
These efforts suggest that HD 63935 b will have the most clearly visible atmosphere of its class. 
It is the best target for transmission spectroscopy (ranked by Transmission Spectroscopy Metric, a proxy for atmospheric observability) in the so-far uncharacterized parameter space comprising sub-Neptune-sized (2.6 $R_\oplus$ $<$ $R_p$ $<$ 4 $R_\oplus$), moderately-irradiated (100 $F_\oplus$ $<$ $F_p$ $<$ 1000 $F_\oplus$) planets around G-stars.
Planet c is also a viable target for transmission spectroscopy, and given the indistinguishable masses and radii of the two planets, the system serves as a natural laboratory for examining the processes that shape the evolution of sub-Neptune planets.

\end{abstract}



\section{Introduction}


Extremely rapid growth in the number of known transiting planets, thanks in large part to the Kepler and K2 missions \citep[e.g.][]{Batalha2013, Thompson2018, Crossfield2016, Kruse_2019}, has enabled population-level studies based on bulk properties like orbital period and size. One of the striking features of the population is that the mode of the  radius distribution lies near 2.5\rearth -- a class of planets lacking in the architecture of our own solar system. These super-earths and/or mini-neptunes bridge the size domain of terrestrials and ice giants in our solar system.

Occurrence rate distributions of these ``bridge'' planets exhibit additional features in the period-radius plane. For example, a feature of particular relevance to this work is the so-called ``radius cliff'', a steep drop in planet occurrence between 2.5 and 4.0 R$_\oplus$ \citep{Borucki2011, Howard2012, Fulton2017} for planets interior to 1AU. Both planets described in this paper fall in this radius regime.
As well, the so-called ``super-Earth" (1-1.8 R$_\oplus$) and ``sub-Neptune" (1.8-4 R$_\oplus$) exoplanets are the modes of the known planet radius distribution within these bridge planets \citep{Fulton2017}, a conclusion supported by modelling \citep[e.g.][]{Owen_2017}.  
Description of the population in relation to the radius ``cliff'' and ``valley'' has become ubiquitous since their discovery. 

As theoretical understandings have matured with acquisition of mass information, a variety of origins for the radius cliff have been proposed.
\cite{Kite2019} propose atmospheric sequestration into magma as the cause, as larger atmospheres achieve the critical base pressure necessary to dissolve H$_2$ from the atmosphere into the core.
Work is still ongoing to understand how atmospheric observables vary across these features in order to better grasp their underlying physics.
Atmospheric characterization of the planets described in this paper could help provide support for theories of the causes of the radius cliff.

Though confirmed multi-planet systems are only a subset of the full planet sample, substantial work has been done to understand the properties of such systems. \cite{Weiss_2018} identify the ``peas in a pod'' phenomenon in the Kepler sample, which describes the fact that planets of a given size are more likely to have neighbors of a similar size than a random size. The nearly-identical planets in the HD 63935 system conform to this trend. They also identify a trend towards denser inner planets, which may be related to photoevaporation. 

Mass measurements are resource intensive to obtain, in particular for the dim host stars common in the Kepler sample, but they are crucial to understanding the population in detail. Early mass measurements by \cite{Weiss2014} demonstrated that Kepler planets above $\sim$1.8\rearth whose stars were sufficiently bright for RV followup appear to retain substantial H/He envelopes. This is in agreement with theoretical predictions for that size regime \citep{Lopez2014}. 
There has since been a great deal of additional work establishing relations between planetary mass and radius, though for any given planetary radius there is a wide spread in masses, implying substantial compositional diversity \citep[e.g.][]{Rogers_2015, Wolfgang2016, ChenKippingMR2017, Zeng2019Composition}.

Efforts to understand the sub-Neptune population are being supported by the \textit{Transiting Exoplanet Survey Satellite} (TESS) mission \citep{Ricker2014tess}. 
Because \tess is observing bright, nearby stars, among the planets it identifies we expect many high-quality atmospheric targets. One of the \tess Level 1 science goals is to ensure that the masses of 50 planets with radii less than 4 \rearth are measured.
In doing so, we will begin to better understand the processes shaping the underlying distributions of exoplanets. The first TESS catalog paper was recently released, documenting 2241 transiting candidates from the mission \citep{guerrero2021toicatalog}. Ongoing followup work, including that undertaken by our group, the \tess-Keck Survey, continues to release well-determined masses for \tess planets.

The TESS-Keck Survey (TKS) is a consortium performing precise radial velocity followup of TESS planet candidates \citep{Dalba2020TKSI, Weiss2021TKSII, Dai2020TKSIII, Rubenzahl2021TKSIV}. One of our group's primary science goals is measuring a diverse set of planet masses at high enough precision to be suitable for atmospheric characterization \citep{Batalha2019}, in particular with \jwst, which is what led us to observe HD 63935. 
One relevant axis of diversity is in host star spectral type. 
So far, only one sub-Neptune sized planet around a G star has been the subject of an atmospheric characterization study (HD 3167 c - \cite{mikalevans2020}). 
Because of their relatively small transit signals, such planets represent more challenging targets than similar-sized targets around M or K dwarfs.
However, the brightness of the host stars can compensate for this, and a number of compelling targets for atmospheric characterization with JWST around G star hosts have already emerged from TESS \citep[e.g.][Lubin et. al.; in prep, Turtelboom et. al., in prep]{Gandolfi_2018, Mann_2020, weiss2020}.

HD 63935 b, the subject of this paper, is one of the planets we identified as a compelling atmospheric target around a G-type host star.
HD 63935 is a bright (V$_{mag}=8.58$) G5 star at a distance of 49 pc. 
Study of planets around this class of star is valuable to help understand how differences in host star characteristics shape planet formation.
With atmospheric data, we will be able to test theoretical predictions like those of \cite{Lopez2014}, that H/He mass fraction is primarily a function of radius. 
Characterizing these planets will also be valuable for comparative studies with our own solar system. 
Although we have not discovered any planetary systems that closely resemble our own, the reasons for this are likely observational \citep{Martin2015}.
A better understanding of what makes our solar system unique (or not) is important to the search for life. 

In this paper, we present the confirmation of the sub-Neptune planets HD 63935 b and c. Planet b is uniquely well-suited to atmospheric characterization, being the second-best target on the radius cliff and the best in its niche of sub-Neptune-sized (2.6 $R_\oplus$ $<$ $R_p$ $<$ 4 $R_\oplus$), moderately-irradiated (100 $F_\oplus$ $<$ $F_p$ $<$ 1000 $F_\oplus$) planets around G-stars. Planet c is also amenable to atmospheric characterization. We also discuss evidence for a longer-period planetary mass companion to the two confirmed planets, though we ultimately adopt a two-planet model. In Section \ref{sec:selectionalgorithm}, we describe the selection algorithm that our consortium's atmospheres working group uses to select high-quality targets. In Section \ref{sec:stellarcharacterization} we provide a description of our efforts to characterize the planets' host star. In Section \ref{sec:obs}, we will describe the observations we undertook to confirm this planetary system. In Section \ref{sec:analysis}, we will provide details of our analysis and results, and in Section \ref{sec:discussion} discuss their implications.

\section{The TESS-Keck Survey: Atmospheric Target Selection}
\label{sec:selectionalgorithm}

This work is based on data obtained as part of the TESS-Keck Survey (TKS), which performs precise radial velocity (PRV) follow-up of TESS planet candidates using the Keck telescope on Mauna Kea and the APF telescope at Lick Observatory. As part of TKS, we are interested in selecting the TESS planet candidates that would represent the best prospects for atmospheric characterization. TESS has produced (and continues to produce) far too many promising atmospheric targets for one consortium to follow up. Consequently, we attempted to develop an algorithm to prioritize targets to add to our PRV prioritized observing list; see Chontos et. al. (in prep.) for more details about the general selection of TKS targets. 

Our algorithm aims to find high quality atmospheric targets in regions of parameter space mostly bereft of them. We select mostly planets in the sub-Neptune regime, as many highly-observable giant planets are already known and terrestrial planets are, with a few exceptions, not accessible with JWST. As a quantification of ``under-populated parameter space,'' our selection algorithm bins planets in stellar effective temperature, planet radius, and insolation flux. We then select targets that stand out in bins without any characterized planets. A detailed explanation of the algorithm follows. 

The inputs to our algorithm are the star and planet properties from NASA Exoplanet Archive\footnote{\url{https://exoplanetarchive.ipac.caltech.edu/}} and the TESS TOI list\footnote{\url{https://tev.mit.edu/data/collection/193}}. We subject these lists to certain cuts as well as manual inspections of the data validation reports \citep{Twicken2018}. We exclude TOIs with declination $<-20^{\circ}$ or V$\textrm{mag}>12$ for visibility at our facilities and to ensure acceptable SNR, respectively. We also cut planets with R$_p > 10$ R$_\oplus$, as the Jovian population is already reasonably well-sampled (for the atmospheres science case only - TKS as a whole does follow up some large planet candidates). We also exclude stars with T$_{\textrm{eff}} >$ 6500 K. 

After our initial culling of the sample, we calculate an estimated mass for each TESS planet candidate based on the fitting formulae of \cite{ChenKippingMR2017} and \cite{Weiss2014}. With that, we calculate a Transmission Spectroscopy Metric (TSM) value \citep{Kempton2018}. This value is an estimate for the Signal-to-Noise Ratio (SNR) of a planet candidate's atmosphere as observed with the NIRISS instrument on the James Webb Space Telescope. For our purposes, it serves as a proxy for relative observability.

\begin{figure}[h]
    \centering
    \includegraphics[width=0.46\textwidth]{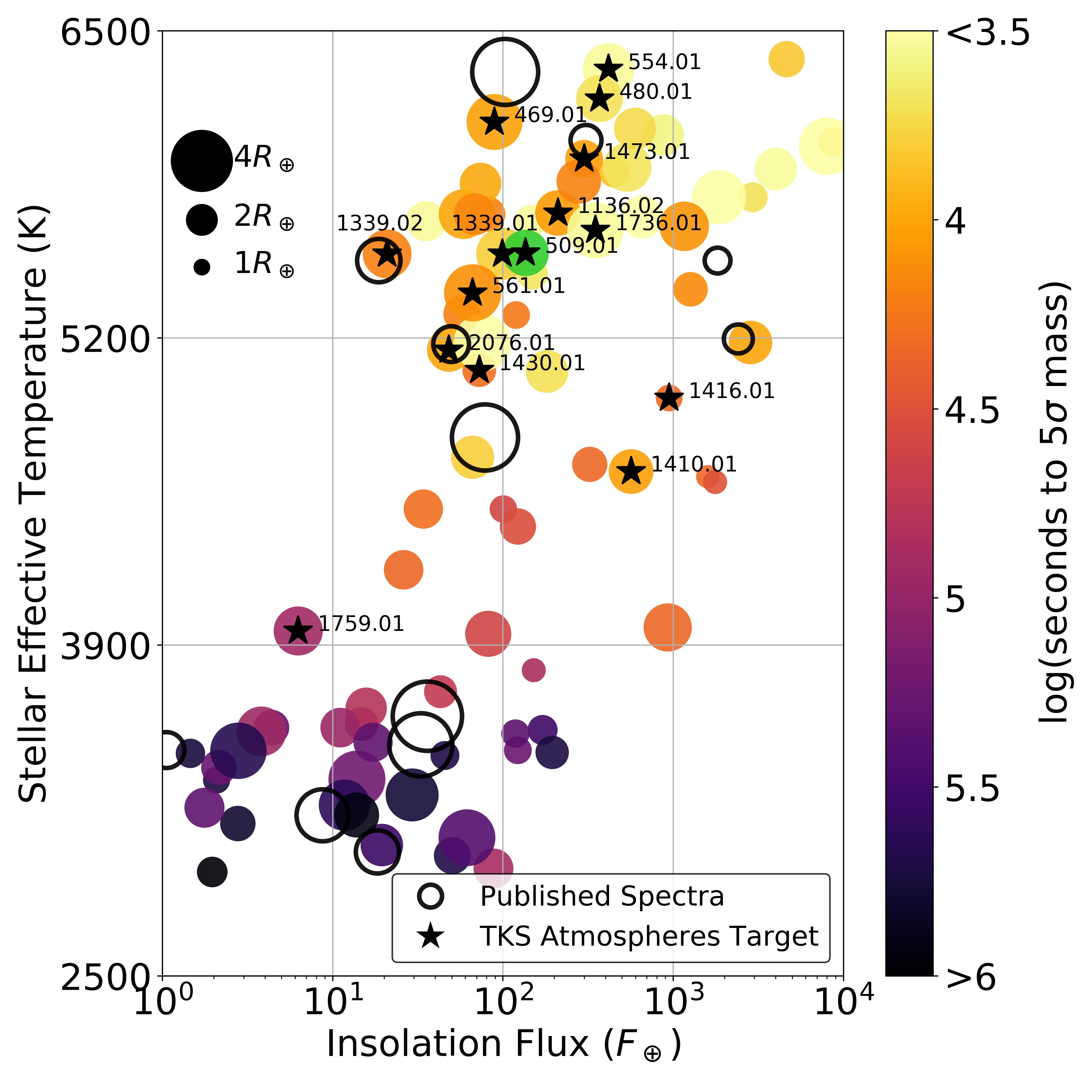}
    \caption{Sub-Neptune TESS planet candidates with estimated TSM values higher than 84 (the value suggested by Kempton et. al. as the cutoff for followup in this size regime), colored by their estimated time to a mass precision of 20\% with HIRES. The figure also plots known planets with published atmospheric transmission spectra observations (most from Hubble's Wide Field Camera 3) as open black circles. Population-level studies are more attainable for the sub-Neptunes with G star hosts, since we can obtain high-precision mass measurements on such planets quickly. TOIs included in the TKS atmospheric sample (sub-Neptune sized planets only) are identified with stars and labelled with their TOI numbers, highlighting the sample's focus on G type host stars. HD 63935 b is colored green. HD 63935 c is not shown as it was not known as a planet candidate when our target list was finalized.} \label{fig:teff-insolation-sample}
\end{figure}

Our selection algorithm then computes a final parameter, which is equal to the TSM value normalized by the expected exposure time on HIRES that would be required to obtain a mass precision better than 20\%, estimated based on \cite{plavchan2015}. This metric was chosen as our ranking parameter in order to select a reasonably large sample of planets. Ranking by TSM alone results in HIRES time spent on M stars, which is sub-optimal for a visual light spectrograph. Cool stars are better suited to characterization by instruments at other facilities like MAROON-X \citep{seifahrt2018maroonx} or CARMENES \citep{quirrenbach2020carmenes}, which have more sensitivity in the red. See fig. \ref{fig:teff-insolation-sample}, which plots the most promising atmospheric targets from \tess with their estimated time to measure the mass with a precision of 20\%, for a visual representation of this. 

With all relevant metrics calculated, our algorithm divides planets into bins along three axes in parameter space: planet radius, stellar effective temperature, and insolation flux. We use five log-uniform radius bins (which conveniently includes edges at 1.7 R$_\oplus$, approximately the location of the radius gap, and 4 R$_\oplus$, dividing the sub- and super-Neptune populations), five log-uniform insolation flux bins, and three stellar effective temperature bins, resulting in 75 bins total, some of which were unpopulated. TOIs that had higher X metric values than other TOIs and known planets in their bin were prioritized for our PRV observing list. There were enough planet candidates meeting this criteria that we could not observe them all, and as a result we focused on those with the highest X metrics. Our algorithm permitted the removal of candidates deemed observationally unsuitable, e.g. by spectral signatures suggestive of an eclipsing binary, or by substantial stellar activity making continued observations infeasible. This selection process is what led us to observe HD 63935, the subject of this paper, which remains the highest-ranked target in its bin by X metric. 

\section{Host Star Characterization}
\label{sec:stellarcharacterization}

We obtained high resolution spectra of HD 63935 (HIP 38374, TIC 453211454; other aliases in Table \ref{tab:stellar-traits}). We used SpecMatch-Syn \citep{Petigura2017} to obtain stellar effective temperature, $\log g$, and metallicity. We then used these values, combined with luminosity and parallax from Gaia \citep{gaiamission, gaiaEDR3}, as priors to obtain the stellar mass, radius, and age using \texttt{isoclassify} \citep{isoclassify-huber-2017, isoclassify-berger-2020} (Table \ref{tab:planet-params}). \texttt{isoclassify} functions by using the input parameters stated above to fit the star to a curve of constant age (isochrone) in the relevant parameter space, allowing the calculation of a radius, mass, and age value. We add a 4\% and 5\% systematic uncertainty to the stellar radius and mass values, respectively, to account for isochrone grid uncertainty, following \cite{tayar2020guide}. Note that our derived age does not account for these uncertainties. The values we obtain for mass and radius are consistent with those provided by SpecMatch Synth, those derived by the \gaia mission \citep{gaiaEDR3}, and those from our spectral energy distribution fitting. The star is slightly smaller than the Sun (R$_*$=0.959R$_\odot$, M$_*$=0.933R$_\odot$), and its \texttt{isoclassify}-derived age suggests that the system is older as well, at $6.8\pm1.8$ Gyr. This is consistent with our nominally low value of $v \sin i = 0.24^{+1.0}_{-0.24}$ km s$^{-1}$ and low activity indicator $\log$ R'$_{HK}$ = -5.06, suggestive of an old, relatively inactive star. The two sectors of TESS photometry do not provide enough information to obtain a reliable rotation period estimate, though we discuss other methods for obtaining this value in Section \ref{sec:gyrochrone}. 

\begin{table}[h!]
  \begin{center}
    \caption{HD 63935 Identifiers \& Gaia Solution}
    \label{tab:stellar-traits}
    \begin{tabular}{lc}
    \hline
    \hline
    \multicolumn{2}{c}{Aliases}\\
      HIP ID & 38374\\
      TIC ID & 453211454\\
      Tycho ID & 783-536-1\\
      \gaia EDR3 ID & 3145754895088191744\\
    \hline
    \multicolumn{2}{c}{\gaia 6D Solution}\\
    Right Ascension & 07:51:42.04\\
    Declination & +09:23:11.40\\
    Parallax & 20.470 $\pm$ 0.019 mas\\
    RA Proper Motion & -78.696 $\pm$ 0.022 mas yr$^{-1}$\\
    Dec Proper Motion & -188.512 $\pm$ 0.013  mas yr$^{-1}$\\
    Radial Velocity & -20.34 $\pm$ 0.19 km s$^{-1}$\\
    \hline

      \end{tabular}
  \end{center}
\end{table}

\subsection{Spectral Energy Distribution and Activity}
\label{sec:sed}
\begin{figure}[!ht]
    \centering
    \includegraphics[width=0.95\linewidth]{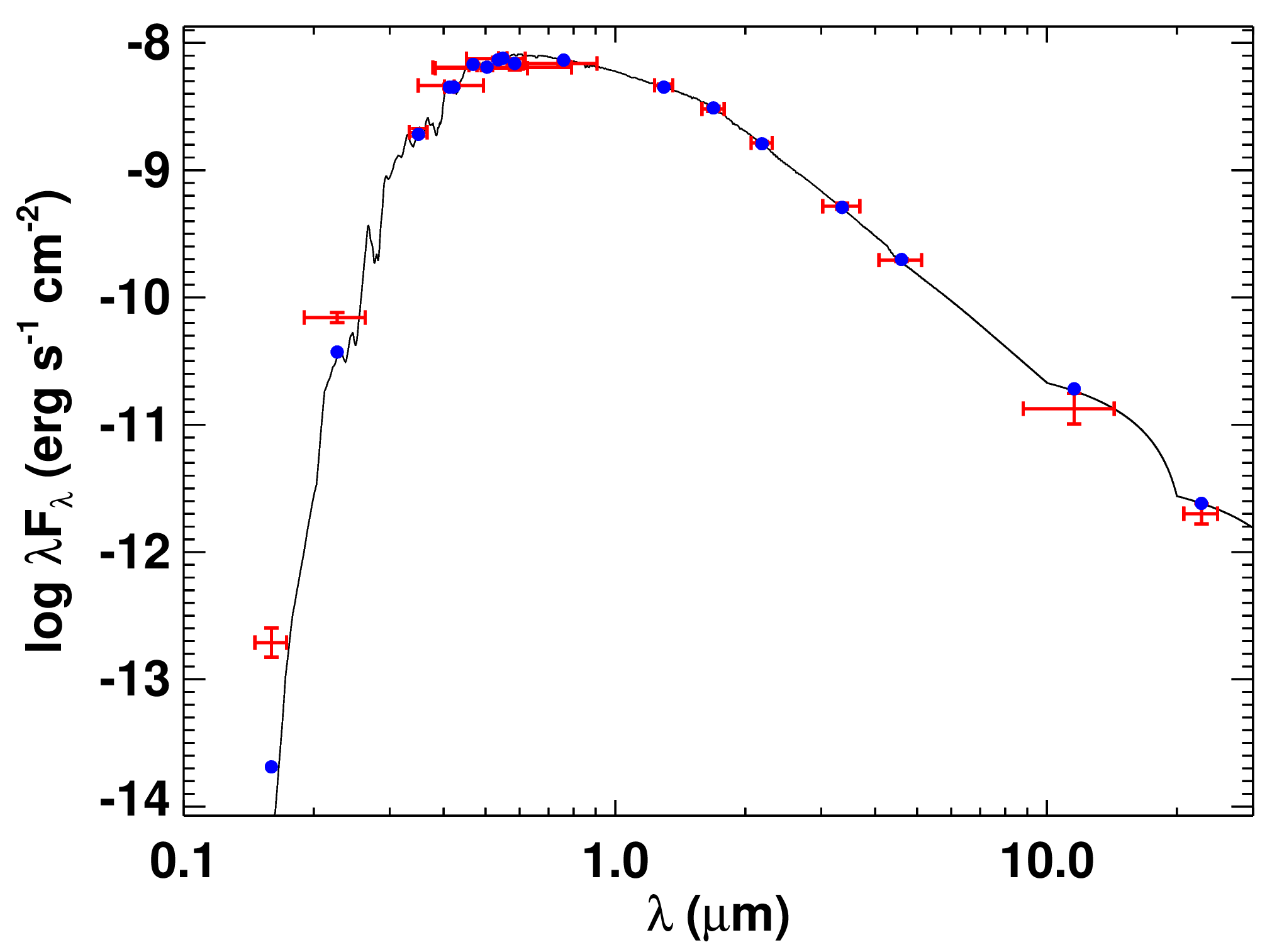}
    \caption{Spectral energy distribution of HD 63935. Red symbols represent the observed photometric measurements, where the horizontal bars represent the effective width of the passband. Blue symbols are the model fluxes from the best-fit Kurucz atmosphere model (black).}
    \label{fig:sed}
\end{figure}

As an independent determination of the stellar parameters, we also performed an analysis of the broadband spectral energy distribution (SED) of the star together with the {\it Gaia\/} EDR3 parallaxes, in order to determine an empirical measurement of the stellar radius, following the procedures described in \citet{Stassun:2016,Stassun:2017,Stassun:2018}. We obtained the $B_T V_T$ magnitudes from {\it Tycho-2}, the Str\"omgren $uvby$ magnitudes from \citet{Paunzen2015}, the $JHK_S$ magnitudes from {\it 2MASS}, the W1--W4 magnitudes from {\it WISE}, the $G G_{\rm BP} G_{\rm RP}$ magnitudes from {\it Gaia}, and the FUV and NUV magnitudes from {\it GALEX}. Together, the available photometry spans the full stellar SED over the wavelength range 0.15--22~$\mu$m (see Figure~\ref{fig:sed}). 

We performed a fit using Kurucz stellar atmosphere models, with the effective temperature ($T_{\rm eff}$), metallicity ([Fe/H]), and surface gravity ($\log g$) adopted from the spectroscopic analysis. The only additional free parameter is the extinction ($A_V$), which we restricted to the maximum line-of-sight value from the dust maps of \citet{Schlegel:1998}. The resulting fit is very good (Figure~\ref{fig:sed}) with a reduced $\chi^2$ of 1.1 and best-fit $A_V = 0.02 \pm 0.02$. Integrating the (unreddened) model SED gives the bolometric flux at Earth, $F_{\rm bol} = 1.060 \pm 0.012 \times 10^{-8}$ erg~s$^{-1}$~cm$^{-2}$3. Taking the $F_{\rm bol}$ and $T_{\rm eff}$ together with the {\it Gaia\/} EDR3 parallax, gives the stellar radius, $R_\star = 0.967 \pm 0.035 R_\odot$. In addition, we can use the $R_\star$ together with the spectroscopic $\log g$ to obtain an empirical mass estimate of $M_\star = 0.82 \pm 0.20 M_\odot$, which is consistent with that obtained via empirical relations of \citet{Torres:2010}, $M_\star = 1.02 \pm 0.06 M_\odot$. These parameters are also consistent with those we derive from \texttt{isoclassify}. Finally, the $R_\star$ and $M_\star$ together yield a mean stellar density of $\rho_\star = 1.59 \pm 0.20$~g~cm$^{-3}$. 

\subsection{Predicted Rotation Period from Gyrochronology}
\label{sec:gyrochrone}
Before photometry confirmed the existence of HD 63935 c as a transiting planet, we were interested in obtaining the stellar rotation period in order to rule that out as the source of the radial velocity signal at 21 days. Although the TESS Sector 34 photometry has since confirmed that candidate, we include the following gyrochronology analysis in the text as it provides novel information about the host star. We used Markov Chain Monte Carlo (MCMC) within \texttt{kiauhoku} \citep{Claytor_2020} to obtain a posterior probability distribution of stellar parameters for HD 63935. For input we used Gaussian priors based on the spectroscopic effective temperature and metallicity, as well as the \texttt{isoclassify}-derived age. Assuming a gyrochronological model, we were then able to predict the rotation period. We performed MCMC using two different braking laws on the same grid of stellar models: one (``fastlaunch'') uses the magnetic braking law presented by \cite{van_Saders_2013}, while the other (``rocrit'') uses the stalled braking law of \cite{van_Saders_2016}. With the fastlaunch model, we predicted $P_\mathrm{rot} = 32.4 \pm 6.5$ days, while we predicted $P_\mathrm{rot} = 31.1 \pm 4.3$ days using the rocrit model. Both of these are consistent with a star somewhat older than the sun, as implied by our measured R'$_{HK}$ value and derived via \texttt{isoclassify}, as well as our non-detection of a rotation period in the \tess data.

\section{Observations}
\label{sec:obs}

\subsection{TESS Photometry}
\label{sec:tess-photometry}

\begin{figure*}[h]
\includegraphics[width=0.95\textwidth]{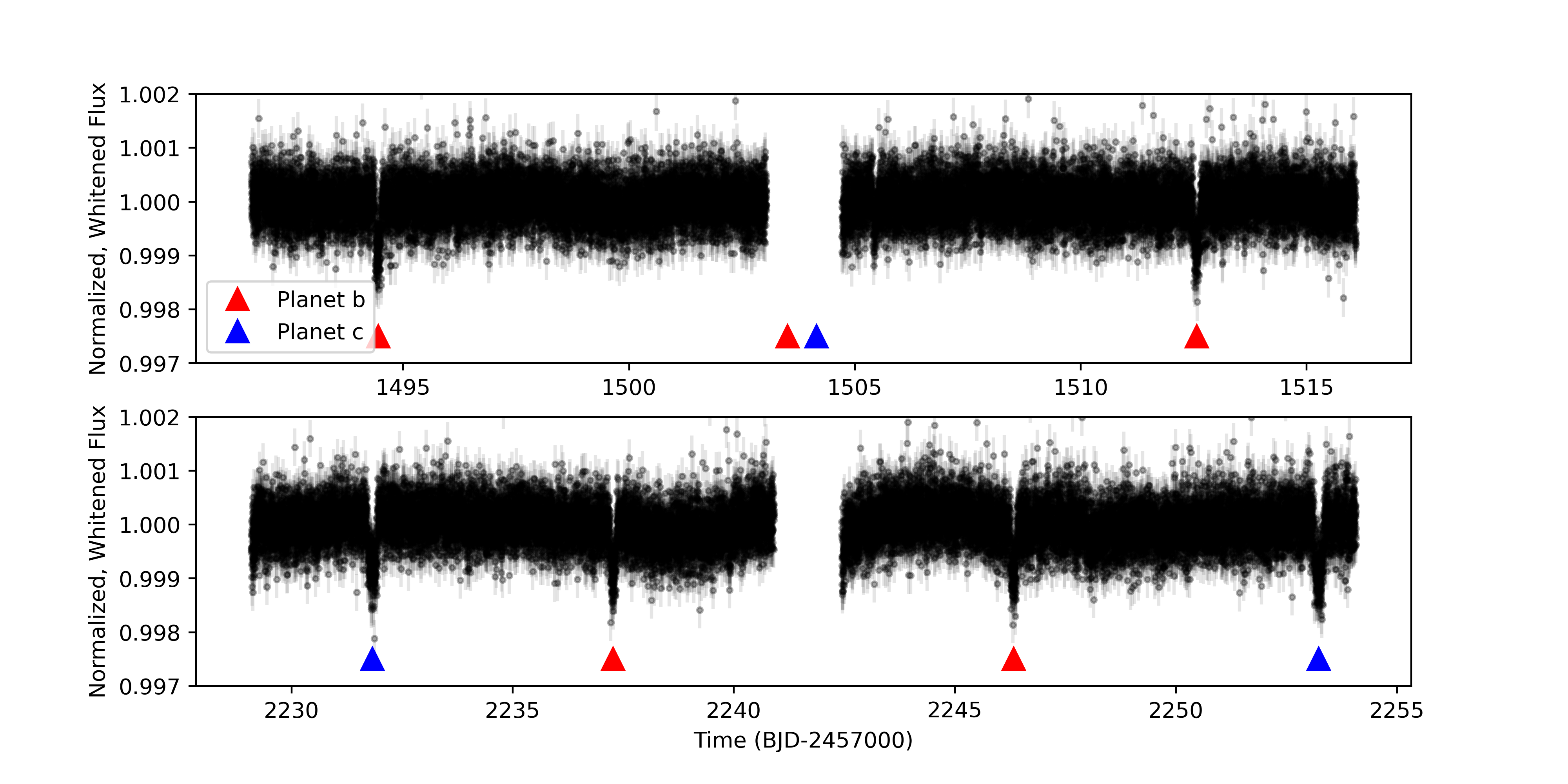}
\caption{The systematics-corrected TESS light curve of HD 63935 in TESS Sectors 7 and 34. Planet b's transits are marked in red and planet c's in blue. Note that the sector 7 data gap contains a transit of both planets. The transit-like event near Time=1506 in Sector 7 is an artifact introduced into the light curve during background subtraction. \label{fig:lc_transit-in-gap}}
\end{figure*}

\begin{figure*}[h]
\includegraphics[width=0.95\textwidth]{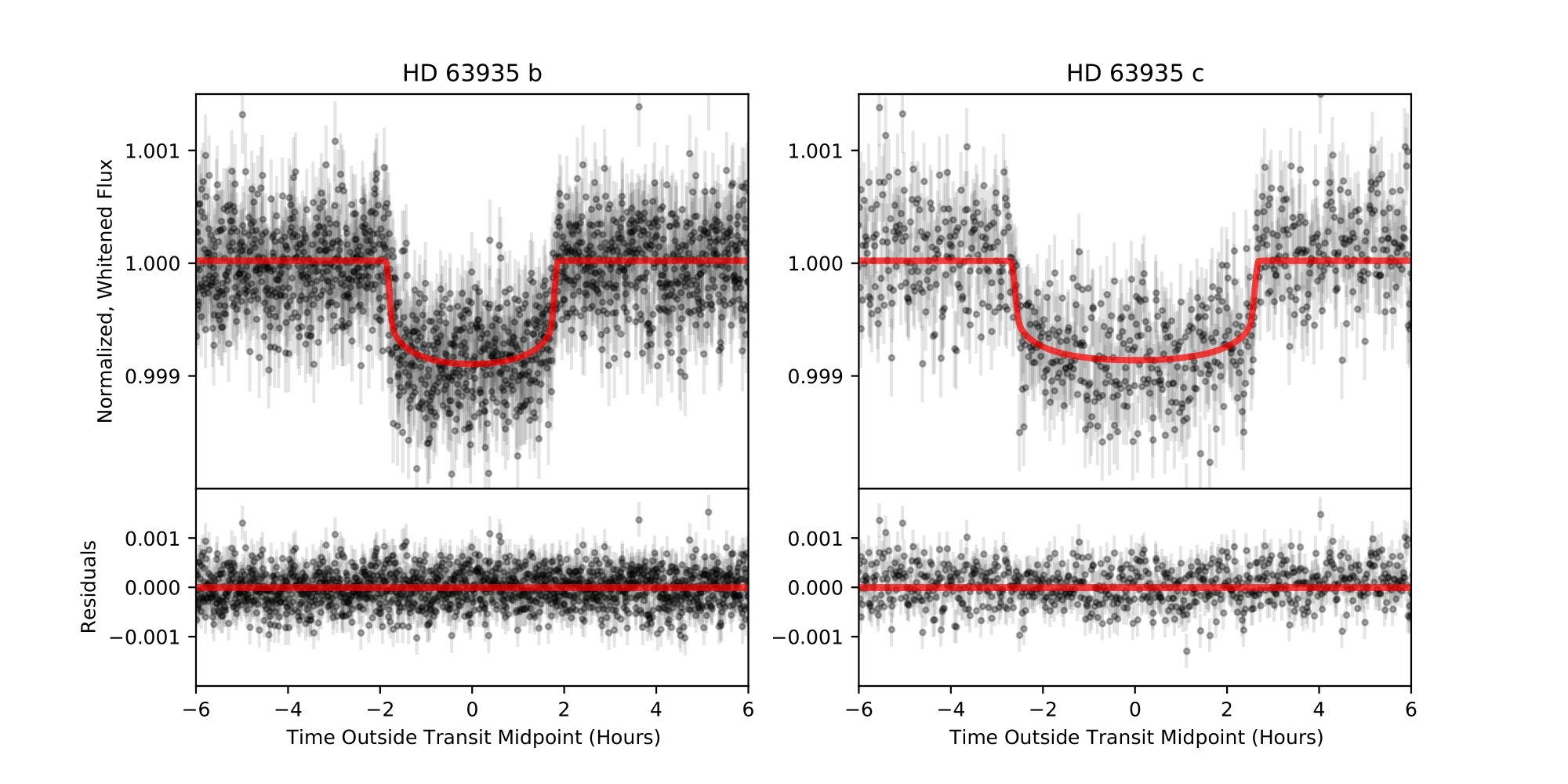}
\caption{Phase-folded transits of HD 63935 b and c. The normalized PDSCAP flux is shown in grey. Our best fit transit model is shown in red. Residuals from each fit are plotted in the lower panels.} \label{fig:phase-folded-transits}
\end{figure*}

HD 63935 (TIC 453211454, TOI 509) was observed by the TESS mission in sector 7 between UT 2019 January 7 and 2019 February 2, and sector 34 between UT 2021 January 13 and 2021 February 9. 
The star was imaged by CCD 4 of Camera 1. 
The data consist of 33208 data points with integration times of two minutes each. 
The Science Processing Operations Center \citep{Jenkins2016spoc} processed the data, generated light curves using Simple Aperture Photometry \citep[SAP][]{Twicken2010, Morris2020}, and removed known instrumental systematics using the Presearch Data Conditioning (PDCSAP) algorithm \citep{Smith_2012, Stumpe_2012, Stumpe_2014}. 
The Sector 7 data contained two transits of planet b, and the Sector 34 data contained two additional transits of planet b and two transits of planet c. The transit of planet c that occurred during Sector 7 happened during a gap in the \tess lightcurve (see Section \ref{sec:tessphotanalysis}. 
For the analysis described here, we downloaded the PDCSAP flux data from the publicly-accessible Mikulski Archive for Space Telescopes (MAST)\footnote{\url{https://archive.stsci.edu/tess/}}. The full light curve is plotted in Figure \ref{fig:lc_transit-in-gap} and the phase-folded transits in Figure \ref{fig:phase-folded-transits}.

\subsection{AO Imaging}
As part of our standard process for validating transiting exoplanets to assess the the possible contamination of bound or unbound companions on the derived planetary radii \citep{Ciardi_2015}, we observed TOI-509 with high-resolution near-infrared adaptive optics (AO) imaging at Palomar and Keck Observatories.

The Palomar Observatory observations were made with the PHARO instrument \citep{Hayward_2001} behind the natural guide star AO system P3K \citep{Dekany_2013} on 2019~Apr~18 UT in a standard 5-point quincunx dither pattern with steps of 5\arcsec\ in the narrow-band $Br-\gamma$ filter $(\lambda_o = 2.1686; \Delta\lambda = 0.0326~\mu$m).  Each dither position was observed three times, offset in position from each other by 0.5\arcsec\ for a total of 15 frames; with an integration time of 1.4 seconds per frame, the total on-source time was 21 seconds on target. PHARO has a pixel scale of $0.025\arcsec$ per pixel for a total field of view of $\sim25\arcsec$. These observations were taken at an airmass of 1.1858. 

The Keck Observatory observations were made with the NIRC2 instrument on Keck-II behind the natural guide star AO system \citep{Wizinowich_2000} on 2019-Mar-25 UT in the standard 3-point dither pattern that is used with NIRC2 to avoid the left lower quadrant of the detector which is typically noisier than the other three quadrants. The dither pattern step size was $3\arcsec$ and was repeated twice, with each dither offset from the previous dither by $0.5\arcsec$.  NIRC2 was used in the narrow-angle mode with a full field of view of $\sim10\arcsec$ and a pixel scale of approximately $0.0099442\arcsec$ per pixel.  The Keck observations were made in the narrow-band $Br-\gamma$ filter $(\lambda_o = 2.1686; \Delta\lambda = 0.0326~\mu$m) with an integration time of 0.5 second for a total of 4.5 seconds on target. Observations were taken in narrow camera mode with a 1024” x 1024” FOV and at an airmass of 1.43.

The AO data were processed and analyzed with a custom set of IDL tools.  The science frames were flat-fielded and sky-subtracted.  The flat fields were generated from a median average of dark subtracted flats taken on-sky.  The flats were normalized such that the median value of the flats is unity.  The sky frames were generated from the median average of the 15 dithered science frames; each science image was then sky-subtracted and flat-fielded.  The reduced science frames were combined into a single combined image using a intra-pixel interpolation that conserves flux, shifts the individual dithered frames by the appropriate fractional pixels, and median-coadds the frames (Figure \ref{fig:ao_fullfov}).  The final resolution of the combined dithers was determined from the full-width half-maximum of the point spread function; 0.094\arcsec\ and 0.050\arcsec\ for the Palomar and Keck observations respectively.

\subsection{Ground-Based Photometry}

The TESS pixel scale is $\sim 21\arcsec$ pixel$^{-1}$, and photometric apertures typically extend out to roughly 1 arcminute, which generally results in multiple stars blending in the TESS aperture. An eclipsing binary in one of the nearby blended stars could mimic a transit-like event in the large TESS aperture. We conducted ground-based photometric follow-up observations as part of the TESS Follow-up Observing Program (TFOP)\footnote{https://tess.mit.edu/followup} with much higher spatial resolution to confirm that the transit signal of HD 63935 b is occurring on-target, or in a star so close to HD 63935 that it was not detected by Gaia DR2.

\subsubsection{MuSCAT}

We observed one partial transit of HD 63935 b on 2019 March 24 from 10:29 to 15:09 in UTC covering the expected egress, with the multi-color simultaneous camera MuSCAT \citep{Narita2015}, which is mounted on the 1.88 m telescope of the Okayama Astronomical Observatory in Okayama, Japan. MuSCAT has three optical channels each equipped with a 1024x1024 pixels CCD camera, enabling g-, r-, and z$_s$-band simultaneous imaging. Each camera has a pixel scale of $0 \farcs 358$ per pixel, providing a field of view (FOV) of 6.1×6.1\arcsec. The exposure times were (10, 3, 3) sec for g, r, z$_s$-bands, respectively.
 
We performed standard aperture photometry using the custom photometry pipeline described in detail in \cite{Fukui2011}. The adopted aperture was 20 pixels or 7\arcsec which excludes any nearby stars as the source of the signal of HD 63935 b. Our precision was not enough to detect the expected 0.09\% transit signal on target in each band, but the data ruled out deep eclipses in all nearby stars within the field of view that are consistent with the transit depth from TESS. 

\subsubsection{LCOGT}
\label{sec:lco-obs}

\begin{figure}
    \centering
    \includegraphics[width=0.47\textwidth]{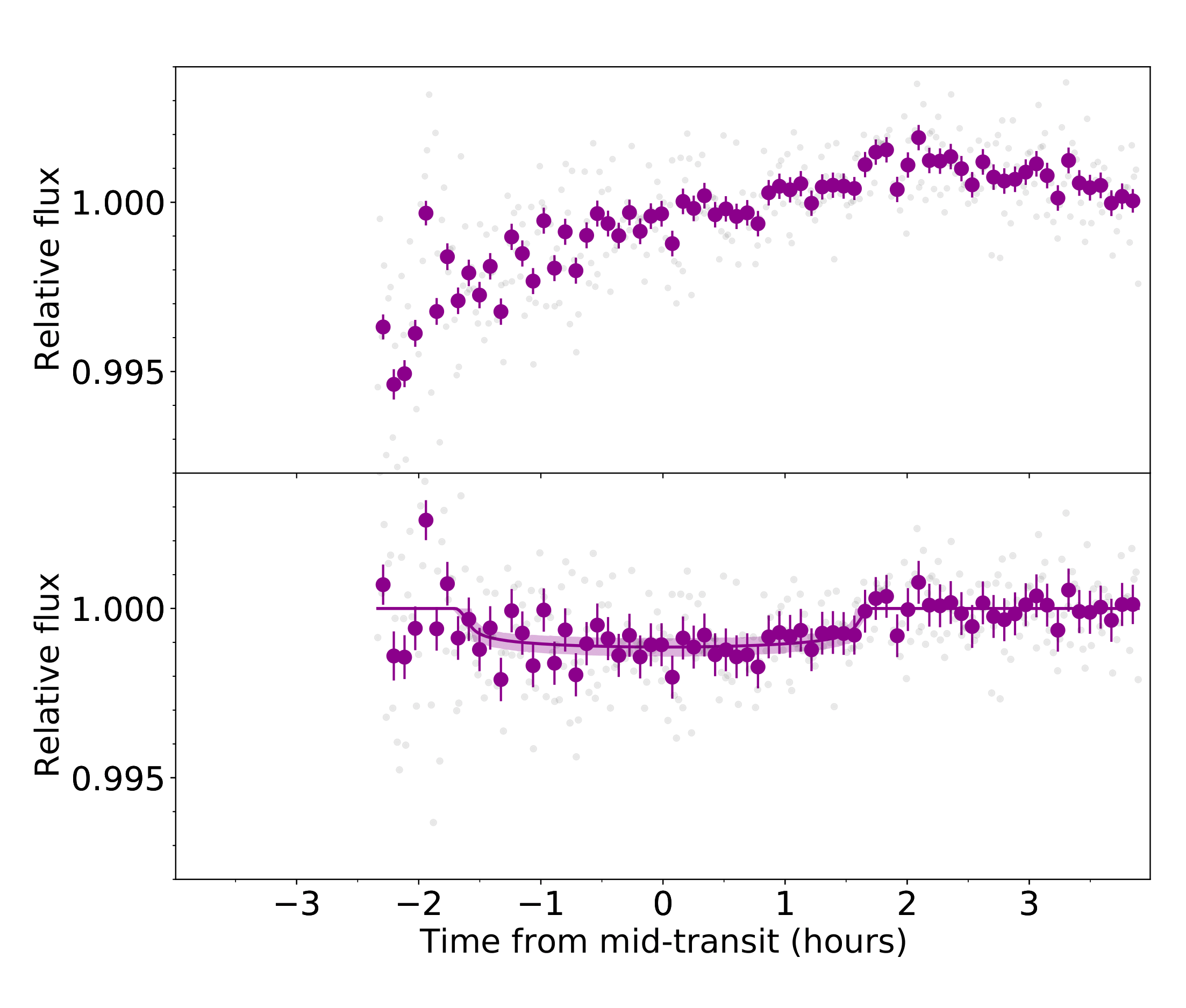}
    \caption{Transit of HD 63935 b observed by the LCOGT 1m telescope at McDonald Observatory on UT 2020-11-20.  Top: Observed differential photometry (grey) with data binned in 5-minute intervals shown in color.  Bottom: Transit model fit, with detrending on airmass and BJD; the shaded region shows the 68.3\% credible interval of the posterior from the MCMC fit (see Sec.\ \ref{sec:lco-fit}). The errorbars in the lower plot include a jitter term determined by the fit. \label{fig:lco-lightcurve}}
\end{figure}

\begin{figure}[h]
    \centering
    \includegraphics[width=0.46\textwidth]{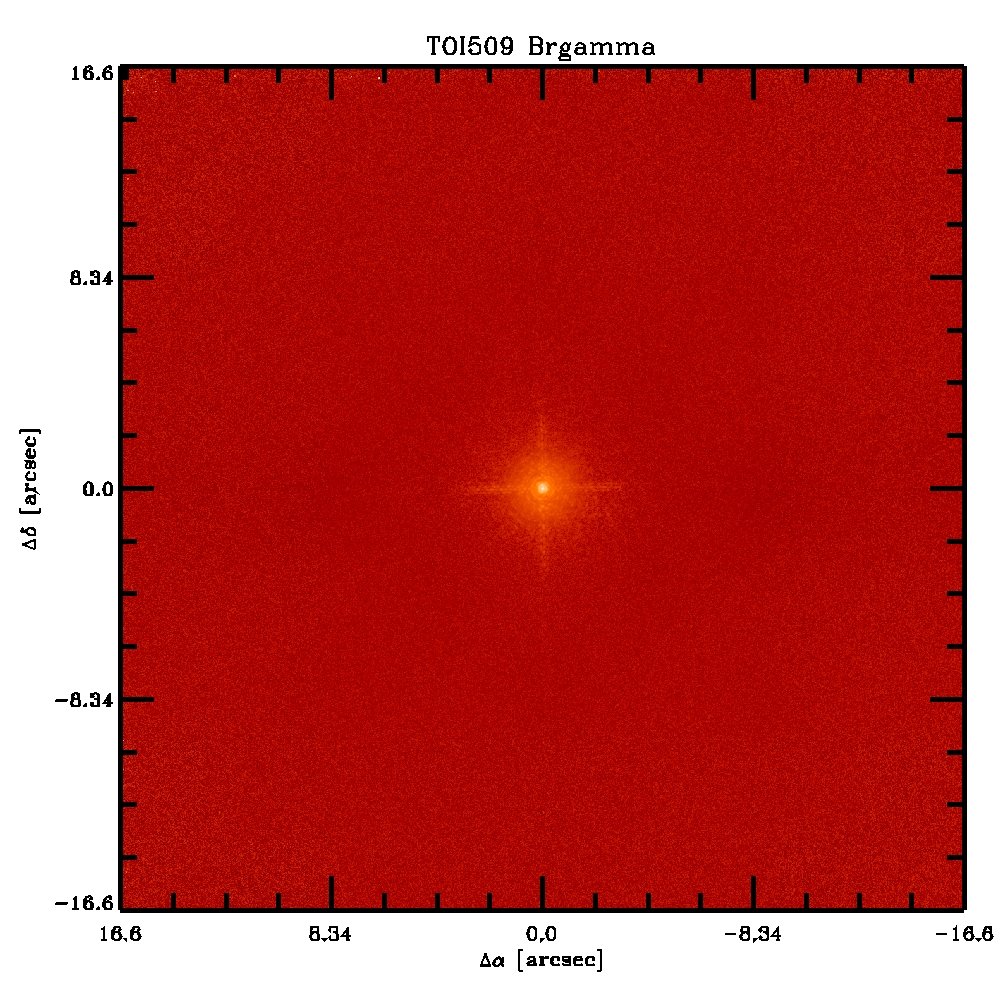}
    \caption{Final combined full field of view dithers of the Palomar observations showing no companions within the TESS pixels.} \label{fig:ao_fullfov}
\end{figure}

We observed a full transit of HD 63935 b in the Pan-STARRS $Y$ filter (central wavelength 1004 nm) on UTC 2020-11-20 from the Las Cumbres Observatory Global Telescope (LCOGT) \citep{Brown2013} 1.0\,m network node at McDonald Observatory. We used the {\tt TESS Transit Finder}, which is a customized version of the {\tt Tapir} software package \citep{Jensen:2013}, to schedule our transit observations. The $4096\times4096$ LCOGT SINISTRO cameras have an image scale of $0\farcs389$ per pixel, resulting in a $26\arcmin\times26\arcmin$ field of view. The images were calibrated by the standard LCOGT {\tt BANZAI} pipeline \citep{McCully:2018}, and photometric data were extracted with {\tt AstroImageJ} \citep{Collins:2017}. The images were focused and have typical stellar point-spread-functions with a full-width-half-maximum (FWHM) of $\sim 1\farcs6$, and circular apertures with radius $\sim 7\farcs 8$ were used to extract the differential photometry. The light curve is presented in Figure \ref{fig:lco-lightcurve}.

\subsection{Ground-Based Spectroscopy}

\subsubsection{LCO NRES}
We obtained a spectrum of the target with the automated LCOGT 1m/NRES optical (380-860) spectrograph \citep{Brown2013, Siverd2018}, in order to characterize the star and look for signs of a stellar binary system. 
The observation was done on UT 2019 March 22, at the McDonald Observatory node of the LCOGT network. 
We observed the target with two consecutive 20 minute exposures that were processed by the LCO Banzai reduction pipeline \citep{McCully:2018} and then stacked together for a spectrum with an effective 40 minutes exposure time and a SNR of 73. 
The reduced spectrum was processed by the SpecMatch-Syn pipeline \citep{Petigura2017} where spectral and stellar parameters were derived while accounting for the target's distance derived from the Gaia DR2 parallax \citep{Gaia18}. 
The spectrum did not show evidence for a second set of lines, and the SpecMatch-Syn analysis showed the target is a slowly rotating ($v\sin i = 0.24^{+1.0}_{-0.24}$ km s$^{-1}$) Sun-like star with an absolute RV ($-20.6 \pm 0.1$ km s$^{-1}$) consistent with the Gaia DR2 RV ($-20.34 \pm 0.19$ km s$^{-1}$).

\subsubsection{TRES}
We obtained two spectra, on UT 2019 March 28 and 2019 April 04, using the Tillinghast Reflector Echelle Spectrograph (TRES) on the 1.5m telescope at the Whipple Observatory on Mt. Hopkins in Arizona. TRES is an optical echelle spectrograph with a wavelength range of $385-910$ nm and has a resolution of $R=44,000$ \citep{gaborthesis, TRES}. The two TRES observations were well-separated in orbital phase (4.21 and 4.59) of the photometric ephemeris and were used to derive relative radial velocities. Using the strongest observed spectrum as a template, the second spectrum was cross-correlated order-by-order in the wavelength range $426-628$ nm. The observed template spectrum was assigned a velocity of zero and the small velocity difference between the two spectra was 13 m/s, which ruled out a stellar or brown-dwarf companion as the source of the transit-like events. The TRES observations also revealed a spectrum very similar to that of the Sun, with line broadening due to rotation of less than 4 km/s and no indication of surface activity, such as emission at Ca II H$\&$K, thus confirming that this target was well suited for PRV work.

The stellar effective temperature ($T_\textrm{eff}$), metallicity ([Fe/H]), surface gravity ($\log g$), and rotation ($v\sin i$) were also determined using the Stellar Parameter Classification tool (SPC) on the TRES spectra \citep{Buchhave2012}. SPC cross-correlates the observed spectra against a library of synthetic spectra calculated using Kurucz model atmospheres \citep{Kurucz} and does a multi-dimensional fit for the stellar parameters that give the highest peak correlation value. These stellar parameter estimates are within 1$\sigma$ agreement with the results from the  SpecMatch-Syn analysis of the PRV observations.

\subsubsection{TNG/HARPS-N}

Between UT 2019 April 2 and 2019 April 29 we collected 11 spectra of TOI-509 with the HARPS-N spectrograph \citep[(383-693 nm, R$\approx$115\,000]{2012SPIE.8446E..1VC} mounted at the 3.58-m Telescopio Nazionale Galileo (TNG) of Roque de los Muchachos Observatory in La Palma, Spain, under the observing programmes CAT19A\_162 (PI: Nowak) and CAT19A\_96 (PI: Pall\'e). 
The exposure time was set to 600--900 seconds, based on weather conditions and scheduling constraints, leading to a SNR per pixel of 59--119 at 5500\,\AA. 
The spectra were extracted using the off-line version of the HARPS-N DRS pipeline \citep{2014SPIE.9147E..8CC}, version 3.7. 

Doppler measurements and spectral activity indicators (Bisector Inverse Slope (BIS), full-width at half maximum (CCF\_FWHM) contrast (CCF\_CTR) of the cross-correlation function (CCF), Mount-Wilson S-index and $\log\mathrm{R_{HK}}$ index) were measured using an on-line version of the DRS, the YABI tool\footnote{Available at \url{http://ia2-harps.oats.inaf.it:8000}.}, by cross-correlating the extracted spectra with a G2 mask \citep{1996A&AS..119..373B}. We measure a $log\textrm{R'HK}$ value of -4.94.
We also used {\tt serval}\footnote{\url{https://github.com/mzechmeister/serval}} \citep{2018A&A...609A..12Z} to measure relative RVs, chromatic index (CRX), differential line width (dLW), and H$\alpha$ index. 
The uncertainties of the relative RVs measured with {\tt serval} are in the range 0.5--1.4$\mathrm{m} \mathrm{s}^{-1}$, with a mean value of 0.83$\mathrm{m} \mathrm{s}^{-1}$.

\subsubsection{Keck HIRES}

Between 2019 August and 2021 March, we obtained 51 high-resolution spectra of HD 63935 with the HIRES instrument \citep{Vogt1994} on the 10m Keck 1 telescope at the W.M. Keck Observatory on Maunakea, Hawai'i. We obtained spectra with the C2 decker, which has dimensions of 14" x 0.86" and spectral resolution R$\approx$60000 at 500 nm. The chosen exposure meter setting regulates S/N at 200 photons per pixel and the resulting median exposure time is 185 s. We obtained radial velocity measurements from the spectra using the method described in \cite{Howard2010}. The RMS value of our radial velocities before fitting for any planets was 4.22 ms$^{-1}$, and the median internal uncertainty was 1.1 ms$^{-1}$. Our measured $\log \textrm{R'HK}$ value is -5.04, indicating a relatively low-activity star and consistent with the value measured by HARPS-N. 

\subsubsection{APF-Levy}

Between 2019 August and 2021 February, we obtained 100 spectra of HD 63935 with the Levy Spectrograph, a high resolution slit-fed optical (500-620nm) echelle spectrograph \citep{Radovan2010APF} on the Automated Planet Finder Telescope (APF) at Lick observatory\citep{Vogt2014}. We observed the star using the W decker, which has dimensions of 3" x 1" and R$\approx$114000 between 374 and 970 nm. Our median exposure time with APF was 1200 seconds. We acquired one or two observations per night. Nightly observations are binned to improve the RV precision. The RMS values of our radial velocities before fitting for any planets was 8.27 ms$^{-1}$, and the median uncertainty was 1.8 ms$^{-1}$. We excluded data points with RV uncertainties $>$5 ms$^{-1}$, which resulted in the removal of five points. All such points were clear outliers and had $<$800 counts on the detector, indicating low data quality.

\begin{table}[h!]
  \begin{center}
    \caption{Radial Velocities}
    \label{tab:RVs}
    \begin{tabular}{lccc}
    \hline
    \hline
      Time (BJD) & RV (m s$^{-1}$) & RV Unc. (m s$^{-1}$) & Inst.\\
      \hline
      2458733.13862 & -6.81 & 1.29 & HIRES\\
      2458744.13959 & -8.04 & 1.11 & HIRES\\
      2458777.06304 & 0.28 & 1.16 & HIRES\\
      2458788.11234 & -11.38 & 1.06 & HIRES\\
      2458795.00493 & -5.68 & 1.00 & HIRES\\
      \hline
      
      \end{tabular}
  \end{center}
        \footnotesize{A sample of the radial velocities, uncertainties, and instruments for our data on HD 63935. The full table of radial velocity data is available online, which includes the Mt. Wilson S-value activity indicators.}
\end{table}


\section{Analysis and Results}
\label{sec:analysis}

\subsection{TESS Photometry Analysis}
\label{sec:tessphotanalysis}

We used \texttt{juliet} \citep{Espinoza2019juliet} to model the light curve data available for HD 63935. \texttt{juliet} serves as a wrapper for a variety of existing publicly-available tools. Our transit modelling used the functions based on \texttt{batman} \citep{Kreidberg2015} for transit fitting and PyMultiNest \citep{Buchner2014} for the sampling of parameter space. We fit for both planets' periods, crossing times, transit depths, and impact parameters, as well as two quadratic limb darkening parameters (following \cite{Kipping2013}, the out-of-transit flux, and a jitter term from TESS. We keep all other parameters fixed, including the mean stellar density, for which we fix the value to that derived in Section \ref{sec:stellarcharacterization}. We impose normal priors centered on the SPOC values for period and crossing time \citep{Jenkins_2002, Jenkins2010, Li_2019}, all with widths of 0.1 days, and constrain the occultation fraction and impact parameter to a uniform range between 0 and 1. 

When we began to study this system, the gap in the TESS Sector 7 photometry (Sector 34 data had not been obtained yet) combined with only two transits meant that two periods ($\sim18$ days, the one initially reported by SPOC, and $\sim$9 days, which we ultimately selected based on RV measurements and was later confirmed by TESS S34) were possible for planet b. The Sector 7 light curve also provided no indication of the existence of planet c, as its transit fell in the data gap as well. We selected the correct period (9 days) and identified planet c as a candidate based on our radial velocity observations (see Section \ref{sec:RVanalysis}). Both of our predictions were validated by the Sector 34 light curve, which confirmed the 9 day period as the correct one for planet b and provided 2 transits of planet c at almost exactly the period predicted by our radial velocity data (21.40 days compared to our predicted value of 21.35 days). Both planets have high SNR transits (18.6 and 23.0 for planets b and c, respectively). The results are of these fits are reported in Table \ref{tab:planet-params}. 

\begin{figure}[h]
    \centering
    \includegraphics[width=0.46\textwidth]{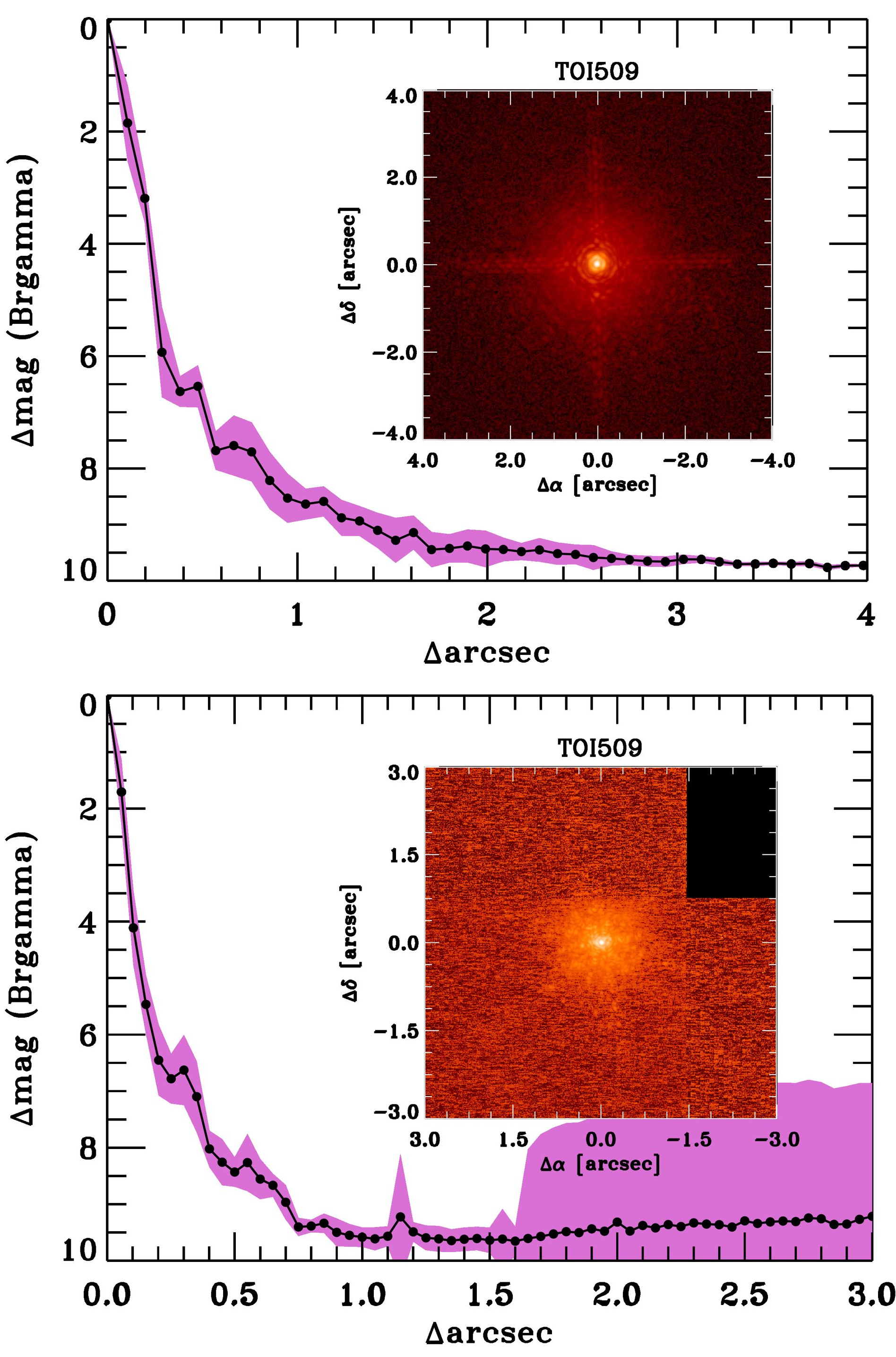}
    \caption{Companion sensitivity for the near-infrared adaptive optics imaging at Palomar (above) and Keck (below).  The black points represent the 5$\sigma$ limits and are separated in steps of 1 FWHM; the purple represents the azimuthal dispersion (1$\sigma$) of the contrast determinations (see text). The inset image is of the primary target showing no additional close-in companions.}\label{fig:ao_contrast}
\end{figure}

\subsubsection{LCO Photometry Analysis}
\label{sec:lco-fit}

As a check on the orbital solution for planet b (in particular the orbit period), we performed an independent fit of the LCO lightcurve (Fig.\ \ref{fig:lco-lightcurve}, Sec.\ \ref{sec:lco-obs}) using \textsf{exoplanet} \citep{exoplanet:exoplanet}.  We fixed $T_0$ at the value found from the TESS lightcurve fit.  The fit to the LCO lightcurve gives 68\% credible intervals on the posterior values found for $P = 9.058875^{+0.000021}_{-0.000034}$, $b = 0.60^{+0.20}_{-0.41}$, $T_d = 3.39 \pm 0.14$ hr, and $R_p/R_* = 0.0325^{+0.0040}_{-0.0043}$. These are consistent with the values from the TESS fit (Table \ref{tab:planet-params}), though the period values are consistent only within the 2-$\sigma$ errors, and not within the 1-$\sigma$ errors.


\subsection{High Resolution Imaging}

To within the limits of the AO observations, no stellar companions were detected. The sensitivities of the final combined AO image were determined by injecting simulated sources azimuthally around the primary target every $20^\circ $ at separations of integer multiples of the central source's FWHM \citep{Furlan_2017, Lund2020}. The brightness of each injected source was scaled until standard aperture photometry detected it with $5\sigma $ significance. The resulting brightness of the injected sources relative to TOI~509 set the contrast limits at that injection location. The final $5\sigma $ limit at each separation was determined from the average of all of the determined limits at that separation and the uncertainty on the limit was set by the rms dispersion of the azimuthal slices at a given radial distance (Figure~\ref{fig:ao_contrast}).

\begin{figure*}
\begin{center}
    \includegraphics[width=0.85\textwidth]{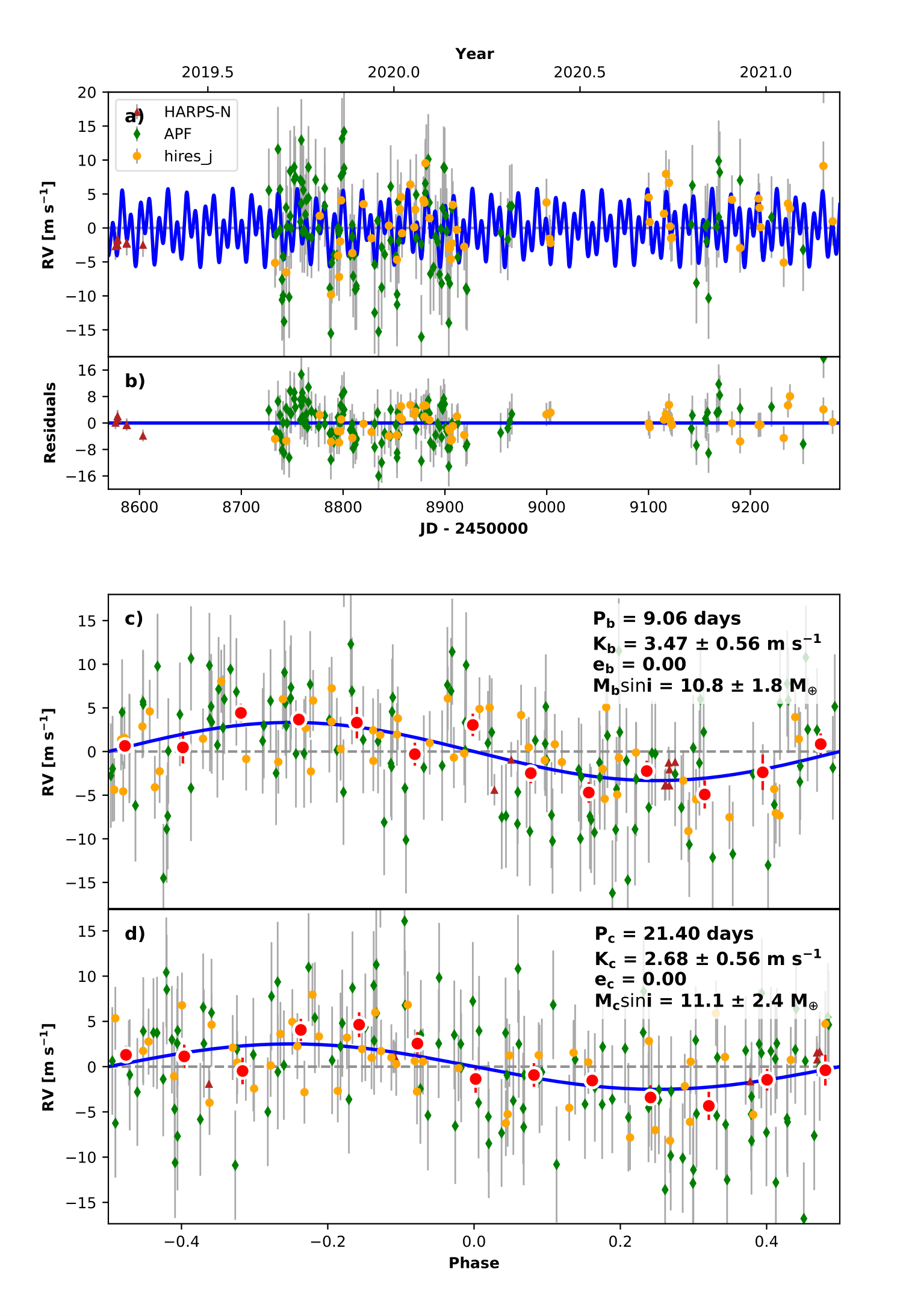}
\end{center}

\caption{Our radial velocity points obtained for HD 63935. a) The complete RV timeseries, including data from HARPS-N (maroon), APF (green), and HIRES (yellow) and our best fit model (blue). b) Our model residuals. Note that the residual appear to exhibit some structure; we discuss this in Section \ref{sec:planet3?} c) Phase-folded radial velocity curve of HD 63935 b. d) Phase-folded radial velocity curve of HD 63935 c. 
\label{fig:phase-folded-rv}}
\end{figure*}

\subsection{Radial Velocity Analysis}
\label{sec:RVanalysis}

We used the \texttt{RadVel}\footnote{\url{https://radvel.readthedocs.io/en/latest/}} package \citep{Fulton2018} to model the radial velocity measurements of HD 63935. \texttt{RadVel} uses the Markov chain Monte Carlo (MCMC) sampler \texttt{emcee} \citep{ForemanMackey2013} to sample the posterior space of the model's parameters. In these fits, we fix the period and time of inferior conjunction to the values derived from photometry (previous section). We enforce circular orbits in our fits. Allowing eccentricity to vary produced a fit which was not preferred by the information criterion analysis. Varying $e_b$ or $e_c$ only were ``somewhat disfavored'' with $\Delta$AICc$=2.28$ and $\Delta$AICc$=3.34$, respectively, while varying both eccentricities was ``strongly disfavored'' with $\Delta$AICc$=6.02$). If the eccentricities are allowed to vary, we find 1$\sigma$ upper limits of 0.16 and 0.29 for planets b and c, respectively. 

Our preliminary single planet fits of this system were dominated by a signal at $\sim$21 days; we identified this signal as corresponding to an additional planet candidate, HD 63935 c, which was confirmed by two transits in TESS Sector 34. The final results of our radial velocity fits are displayed in Figure \ref{fig:phase-folded-rv}.

\begin{table*}[t]
  \begin{center}
    \caption{Complete table of properties used in this analysis}
    \begin{threeparttable}
    \label{tab:planet-params}
    \bgroup
    \def\arraystretch{1}
    \begin{tabular}{llccc} 
        \hline
        \hline
      \multicolumn{1}{l}{\textbf{Parameter}} & 
        \multicolumn{1}{l}{\textbf{Symbol}} &
            \multicolumn{2}{c}{\textbf{Value}} &
                    \multicolumn{1}{c}{\textbf{Units}}\\
      \hline
      \textbf{Stellar Parameters} & & & &\\
      
      Mass\tnote{1} & $M_*$ & \multicolumn{2}{c}{0.933$\pm0.054$} & M$_\odot$\\
      
      Radius\tnote{1} & $R_*$ & \multicolumn{2}{c}{0.959$\pm0.042$} & R$_\odot$\\
      
      Age\tnote{1} & & \multicolumn{2}{c}{6.8$^{+1.8}_{-1.9}$} & Gyr\\
      
      Stellar Effective Temperature\tnote{2} & $T_{\textrm{eff}}$ & \multicolumn{2}{c}{5534$\pm$100}& K \\
      
      Surface Gravity\tnote{2} & $\log g$ & \multicolumn{2}{c}{$4.38\pm$0.1} & cm s$^{-1}$ \\
      
      Metallicity\tnote{1} & [Fe/H] &\multicolumn{2}{c}{$0.07\pm0.06$} & dex\\
      
      Activity Index\tnote{2} & $\log$ R'HK & \multicolumn{2}{c}{-5.06} & \\
      
      V-band Magnitude\tnote{3} & $V_{\textrm{mag}}$ & \multicolumn{2}{c}{8.58} & \\
      
      J-band Magnitude\tnote{3} & $J_{\textrm{mag}}$ & \multicolumn{2}{c}{7.30} & \\
      
      K-band Magnitude\tnote{3} & $K_{\textrm{mag}}$ & \multicolumn{2}{c}{6.88} & \\
      
      Distance\tnote{4} & \textit{d} & \multicolumn{2}{c}{$48.8\pm0.8$} & pc\\
      
      Luminosity\tnote{4} & \textit{L} & \multicolumn{2}{c}{$0.798\pm0.002$} & L$_\odot$\\
      
      Limb Darkening Parameters\tnote{5} & q$_1$, q$_2$ & \multicolumn{2}{c}{$0.27^{+0.28}_{-0.15}$, $0.26^{+0.32}_{-0.18}$} & \\
      
      \hline
      \textbf{Transit Parameters}\tnote{5} & & \textit{Planet b}& \textit{Planet c}& \\
      
      Period & P & $9.058811^{+0.000017}_{-0.000016}$ & $21.4023^{+0.00189}_{-0.00194}$ & days\\
      
      Transit Crossing Time & T$_0$ & 1494.4462$\pm0.0010$ & 2231.8280$\pm0.0014$ & TJD\tnote{6} \\
      
      Occultation Fraction & $R{_p}/R{_*}$ & 0.0285$\pm0.0004$ &  0.0277$\pm0.0004$ & \\
      
      Orbital Separation & a/R$_*$ & 18.64$\pm1.1$ & 33.06$\pm2.0$ & \\
      
      Inclination & $i$ & 88.49$\pm0.0018$ & 88.241$^{+0.0011}_{-0.0013}$ & $^\circ$\\
      
      Transit duration & $T_d$ & 3.36$\pm 0.10$ & 4.85$\pm0.33$ & hours\\
      
      Transit SNR & SNR & 18.6 & 23.0 & \\
      
      \hline
      
      \textbf{RV Parameters}\tnote{7} & & \textit{Planet b} & \textit{Planet c} & \\
      
      Planet Semi-Amplitude & K$_\textrm{amp}$ & $3.18^{+0.55}_{-0.52}$ & $2.71^{+0.52}_{-0.49}$ & m s$^{-1}$\\
      
      
      Eccentricity & $e$ & 0 (fixed) & 0 (fixed) & radians\\
      
      Periastron Passage & $\omega$ & 0 (fixed) & 0 (fixed) & radians\\
      
      \textit{Instrumental and GP Parameters} & & & & \\
      
      
      Linear HIRES Offset & $\gamma_{\rm hires_j}$ & \multicolumn{2}{c}{$3.55^{+0.46}_{-0.40}$} & m s$^{-1}$\\
      
      Linear APF Offset & $\gamma_{\rm apf}$ & \multicolumn{2}{c}{$5.49^{+0.44}_{-0.4}$} & m s$^{-1}$\\
      
      Linear HARPS-N Offset & $\gamma_{\rm HARPS-N}$ & \multicolumn{2}{c}{$0.22^{+3.0}_{-3.1}$} & m s$^{-1}$\\
      
      \hline
      
      \textbf{Derived Parameters} & & \textit{Planet b}& \textit{Planet c}& \\

      Planet Radius & $R_p$ & $\brad\pm\braderr$ & $ \crad\pm\craderr$ & $R_\oplus$\\
      
      Impact Parameter & b & $0.49 \pm 0.02$ & $0.30^{+0.03}_{-0.04}$ & \\
      
      Planet Mass & $M_p$ & $\bmass \pm \bmasserr$ & $\cmass \pm \cmasserr$ & $M_\oplus$\\
      
      Planet Density & $\rho_p$ & $\bdensity \pm \bdensityerr$ & $ \cdensity \pm \cdensityerr$ & g cm$^{-3}$\\
      
      Insolation Flux & $F_p$ & 115.6$\pm4.6$ & 36.7$\pm1.4$ & $S_\oplus$\\
      
      Equilibrium Temperature\tnote{8} & T$_{eq}$ & 911$\pm27$ & 684$\pm21$ & K\\
      
      Transmission Spectroscopy Metric & TSM & \bTSM $\pm$ \bTSMerr& \cTSM $\pm$ \cTSMerr & \\
      
      Emission Spectroscopy Metric & ESM & \bESM $\pm$ \bESMerr & \cESM $\pm$ \cESMerr & \\
     
     \hline
      \end{tabular}
      \begin{tablenotes}
        \item[1] \texttt{isoclassify} \item[2] Specmatch-Syn \item[3] exofop \item[4] \textit{Gaia} \item[5] \texttt{juliet} \item[6] BJD-2457000 \item[7] \texttt{RadVel} \item[8] Assumes zero albedo and full day-night heat redistribution
      \end{tablenotes}
    \egroup
    \end{threeparttable}
  \end{center}
\end{table*}

\subsection{Is There a Third Planet?}
\label{sec:planet3?}
\begin{figure}[H]
\includegraphics[width=0.49\textwidth]{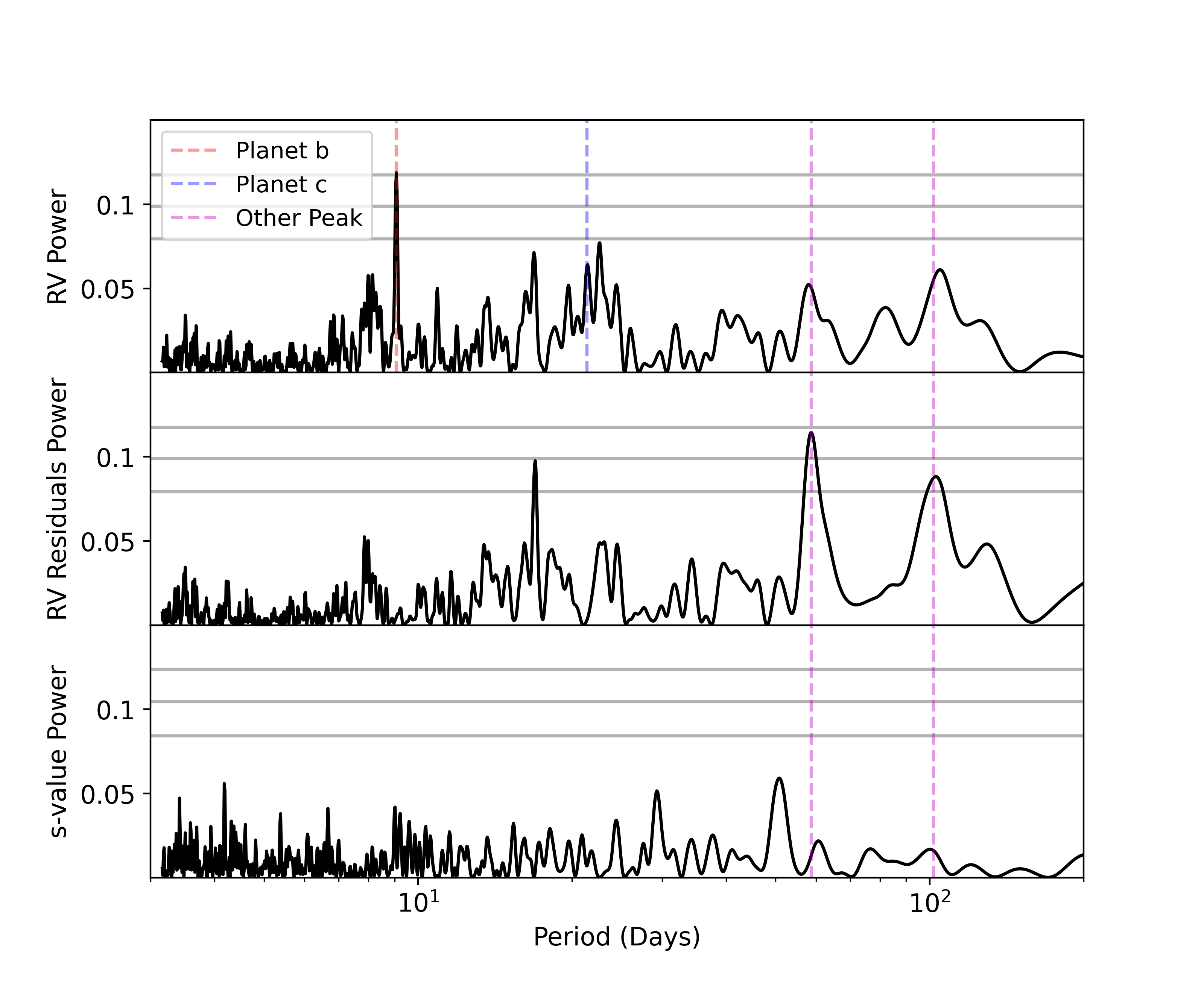}
\caption{Lomb-Scargle periodograms of the radial velocity data for HD 63935, the residuals of the RV data with the signals corresponding to the transiting planets removed, and the Mt. Wilson s-values (activity indicators) for the system. False Alarm Probabilities of 0.1, 0.01, and 0.001 are shown as horizontal grey lines in each plot. The signals corresponding to a potential long-period companion (purple) do not rise to the required level of statistical significance, but also do not correspond to peaks in the s-values. The period near 17 days in the middle panel is ruled out as a transiting planet by the \tess photometry but may be an alias of the stellar rotation period. 
\label{fig:periodogram}}
\end{figure}

The photometry provides clear evidence for two planet candidates, whose presence we confirm with radial velocity followup. The residuals of a two-planet fit, however, show substantial structure. Because the HIRES and APF points in the residuals show similar behavior to each other, instrumental effects are unlikely to be the cause. This suggested that there was something still unaccounted for in our model, which could be a third planet. To test this, we generated a Lomb-Scargle periodogram of the radial velocity residuals from our two-planet fit (Figure \ref{fig:periodogram}). There are two primary visible peaks at longer periods, at $\sim$59 and $\sim$102 days. Because the peaks in the periodograms of the radial velocity residuals and s-values (activity indicators) do not correspond to each other, we consider it unlikely that this signal is caused by stellar activity. The lack of significant periodicity in the S-values suggests no motivation to adopt a Gaussian Process model for our data, consistent with our low value of $\log$R'HK. We have also performed an independent analysis (Section \ref{sec:gyrochrone}) to identify the stellar rotation period, arriving at the conclusion that this period is roughly 30-35 days, and therefore inconsistent with both of the longer-period RV periodogram peaks. 

As a method of examining the significance of these peaks, we perform a bootstrap analysis. In this analysis we repeatedly resample our entire RV dataset with replacement, calculate the power at the locations in period space where the peak is highest in our true dataset, and repeat this 10000 times. Selecting a p value of 0.05, we are unable to reject the null hypothesis for either signal. Therefore, in this paper we have adopted the two-planet model, as the evidence for a third planet does not rise to the level of statistical significance we require. Important to note, however, is that this choice does not substantially impact the mass precision of our two confirmed planets, and the resulting masses are consistent to each other to 1-$\sigma$ for both models. Further radial velocity monitoring or photometric followup could provide clarity as to the source of this additional signal. 

\begin{table}[H]
  \begin{center}
  \caption{Mass Comparison for Different Models}
    \label{tab:diff_model_masses}
    \begin{tabular}{cccc}
    \hline
    \hline
      N$_{planets}$ & $M_b$ (\mearth) & $M_c$ (\mearth) & $M_d \sin i$ (\mearth) \\
      \hline
      2 & $10.8 \pm 1.8$ & $11.1 \pm 2.4$ & n/a\\
      3 ($\textrm{P}_d=58.7$ d) & $11.1^{+1.6}_{-1.5}$ & $12.8^{+2.1}_{-2.0}$ & $16.8^{+2.7}_{-2.6}$\\
      3 ($\textrm{P}_d=101.7$ d) & $9.9^{+1.8}_{-1.7}$ & $11.2^{+2.2}_{-2.0}$ & $20\pm4$\\
      \hline
      
      \end{tabular}
  \end{center}
        \footnotesize{The mass values for planets b and c for different possible models, as well as the $M\sin i$ for a possible third planet where relevant. Note that the masses for the two transiting planets are consistent within 1$\sigma$ no matter which model is selected.}
\end{table}

\section{Discussion}
\label{sec:discussion}

\subsection{Examining Plausible Compositions}
\label{sec:compositions}

In order to better understand this planetary system, we investigated a range of possible compositions based on the planets' bulk density and corresponding positions in mass-radius space. 
This is of particular interest given the substantial degeneracies in composition that exist for planets in this radius range, primarily between ice/volatile dominated planets and rock-dominated interiors with substantial H/He envelopes \citep{Lopez2014}. 
To do this, we used the public tool \texttt{smint}\footnote{\url{https://github.com/cpiaulet/smint}} \citep{Piaulet_2021}, developed by Caroline Piaulet, which utilizes the models of \cite{Lopez2014} and \cite{Zeng2016} to do a Markov Chain Monte Carlo exploration of the posterior space of planetary interior compositions based on their mass, radius, age, and insolation flux. 
We use 25 MCMC walkers and 10000 steps for each in this analysis. 
Our results from \texttt{smint} are that HD 63935 b and c have $3.6\pm0.8$\% and $3.4\pm0.9$\% mass in H/He, respectively.
Both planets could have cores that are intermediate between ice-dominated and rock-dominated. 
However, neither planet is sufficiently dense to be a pure ``water world''; in other words, both are expected to have substantial (few percent mass) H/He envelopes. 
The compositional degeneracies that exist in this part of parameter space are one of the reasons that the sub-Neptune sized planets are so compelling for atmospheric characterization. 
Study of these planets' atmospheres can potentially reveal more about their composition, which in turn can contribute to our understanding of how these planets form and why no analogs exist in our own solar system.  


\begin{figure*}[t]
\includegraphics[width=\textwidth]{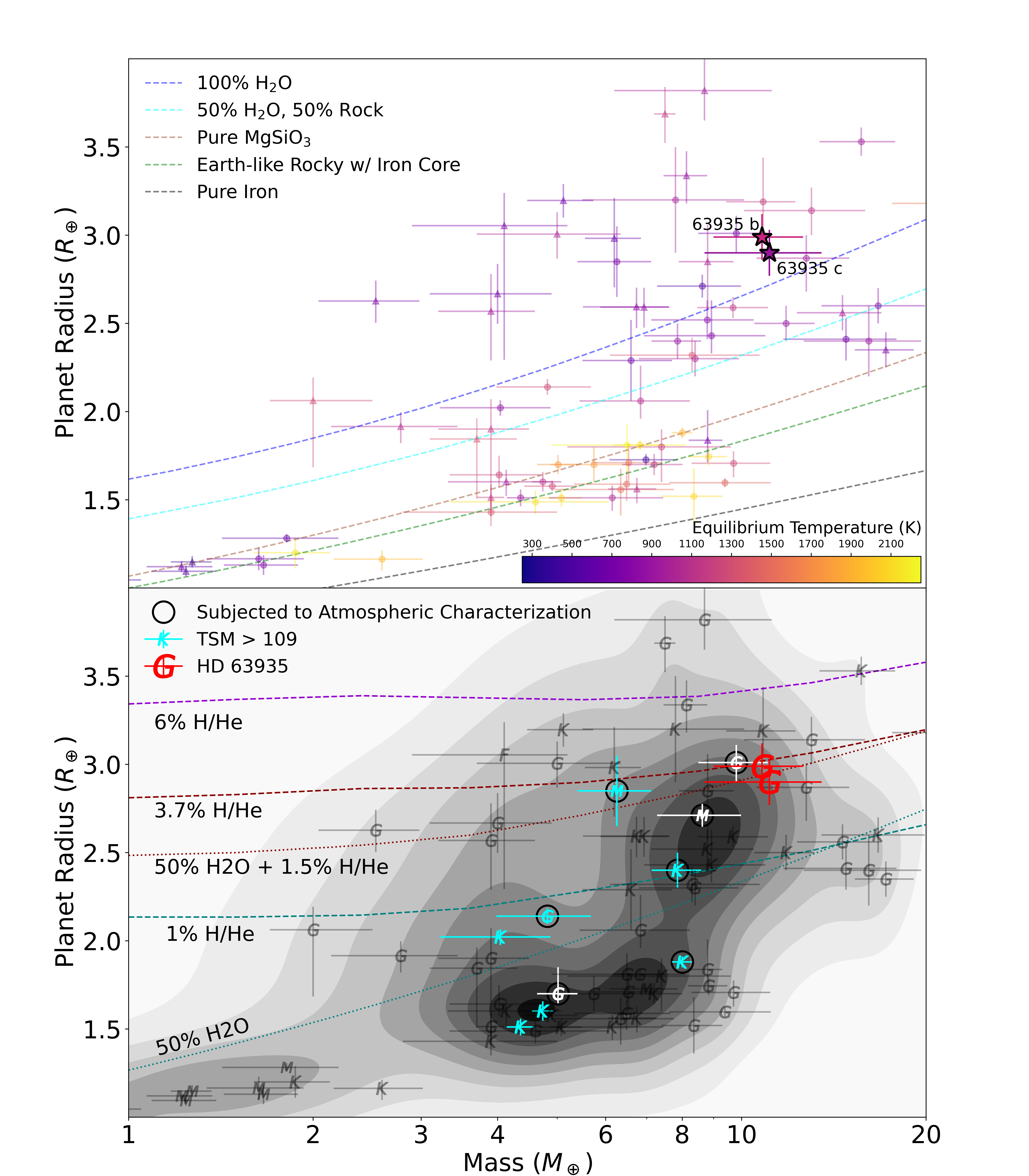}
\caption{The position of HD 63935 b in mass-radius space. Top: the planet sample from \cite{Zeng2019Composition}, colored by equilibrium temperature, as well as a simple subset of composition curves from \cite{Zeng2016} to contextualize the image. Bottom: a density estimation of the same planet sample, with host star spectral type indicated by marker symbol (F, G, K, M). The bimodal distribution of planets around the radius gap at $\sim$1.8 R$_\oplus$ is clearly visible. HD 63935 b and c are emphasized and colored red, with planet b the upper of the two red points. Planets with TSM values higher than HD 63935 b are colored cyan (from top to bottom, GJ 1214 b, HD 97658 b, $\pi$ Mensae c, GJ 9827 d, 55 Cancri e, HD 219134 b, and HD 219134 c), and planets which have been subjected to atmospheric characterization in the past are circled in black. Those which have published atmospheric spectra but lower TSM values (from top to bottom, HD 3167 c, K2-18 b, and HD 3167 b) are colored white. Planets meeting neither criteria are faded grey. The dark red lines correspond to a subset of plausible composition curves for HD 63935 b, from \cite{Lopez2014} and \cite{Zeng2016}. This figure emphasizes the planet's uniqueness as a quality atmospheric target in parameter space.
\label{fig:MR_multipanel}}
\end{figure*}

\subsection{Nearly Twins: How Does HD 63935 Fit in the ``Peas in a Pod'' Structure?} 

\cite{Weiss_2018} identified the phenomenon of Kepler planets in multi-planet systems being more likely to be similar to each other than drawn from a random distribution. They refer to this as ``peas in a pod''.
The two confirmed planets in the HD 63935 system appear to be in line with this trend, as both planets have masses and radii that are consistent to each other within one sigma. Higher precision measurements have the potential to distinguish differences between the two planets. In particular, if planet c is discovered to be higher density (as nominally appears to be the case, though the difference is not presently statistically significant), this would be interesting, because \cite{Weiss2014} note that the larger and lower density planet tends to be the exterior in their sample. They suggest that this result could be explained by photoevaporation. In this case, however, the nominally denser planet is the cooler and less-irradiated one, implying that a different explanation is required. We calculate the $\Lambda$ parameter described by \cite{Fossati_2017}, which estimates whether atmospheric erosion is relevant for a given planet. $\Lambda <25$ at T$_\textrm{eq} = 1000K$ and $\Lambda <35$ at T$_\textrm{eq} = 500K$ (Fig 4 in \cite{Fossati_2017}) are the relevant regions of atmospheric erosion for G star hosts. The respective $\Lambda$ values for HD 63935 b and c are 30 and 42, suggesting that neither planet is likely to experience significant atmospheric Jeans escape. In the absence of atmospheric loss, \cite{Zeng2019Composition} propose that denser outer planets could be explained via impacts of ice-rich planetesimals \citep[see also][]{Marcus2009CollisionalStripping}. 

Also potentially of interest is the differing equilibrium temperatures of these planets ($\sim$911K and $\sim$684K for b and c, respectively). Given their otherwise similar properties, they could serve as an experimental testing ground for the role of insolation on planetary composition and nature. As we describe in more detail in the following section, planets cooler than 1000K are likely to have decreasing spectral feature amplitude in transmission spectra as a result of increased haze formation \citep{Gao_2020}. Observation of such a phenomenon in the HD 63935 system could be used as further evidence for this trend and of particular significance because the planets orbit the same star (eliminating a possible confounding variable). 

\subsection{Assessing Atmospheric Observability}

As described above in Section \ref{sec:selectionalgorithm}, we identified HD 63935 b as the best target for atmospheric characterization followup in its region of parameter space (between 2.6 and 4 Earth radii, between 10 and 100 times Earth insolation flux, and stellar effective temperatures between 5200 and 6500K).
The TSM value for HD 63935 b, incorporating our measured mass, is \bTSM $\pm$ \bTSMerr.
In addition to its uniqueness within our algorithm's defined parameter space bins, planet b is also of interest as a tool to probe the radius cliff, which refers to the apparent drop-off in planet occurrence rate above $\sim 2.5 R_\oplus$. 
In Figure \ref{fig:MR_multipanel}, we identify planets with TSM values equal to or greater than that of HD 63935 b, and find that only one (HD 191939 b, Lubin et. al. (submitted)) falls on the radius cliff.
The physical causes of the radius cliff are hypothesized to be related to atmospheric sequestration \citep{Kite2019}, signs of which may be visible in an atmosphere, further enhancing the target's desirability for characterization. 
Note also that of the planets in that figure, four (HD 219134 b, HD 219134 c, 55 Cnc e, and $\pi$ Mensae c) have host star magnitudes that saturate \jwst, meaning they are not suitable targets for transmission spectroscopy with \jwst. 

\begin{figure*}[t]
\includegraphics[width=0.9\textwidth]{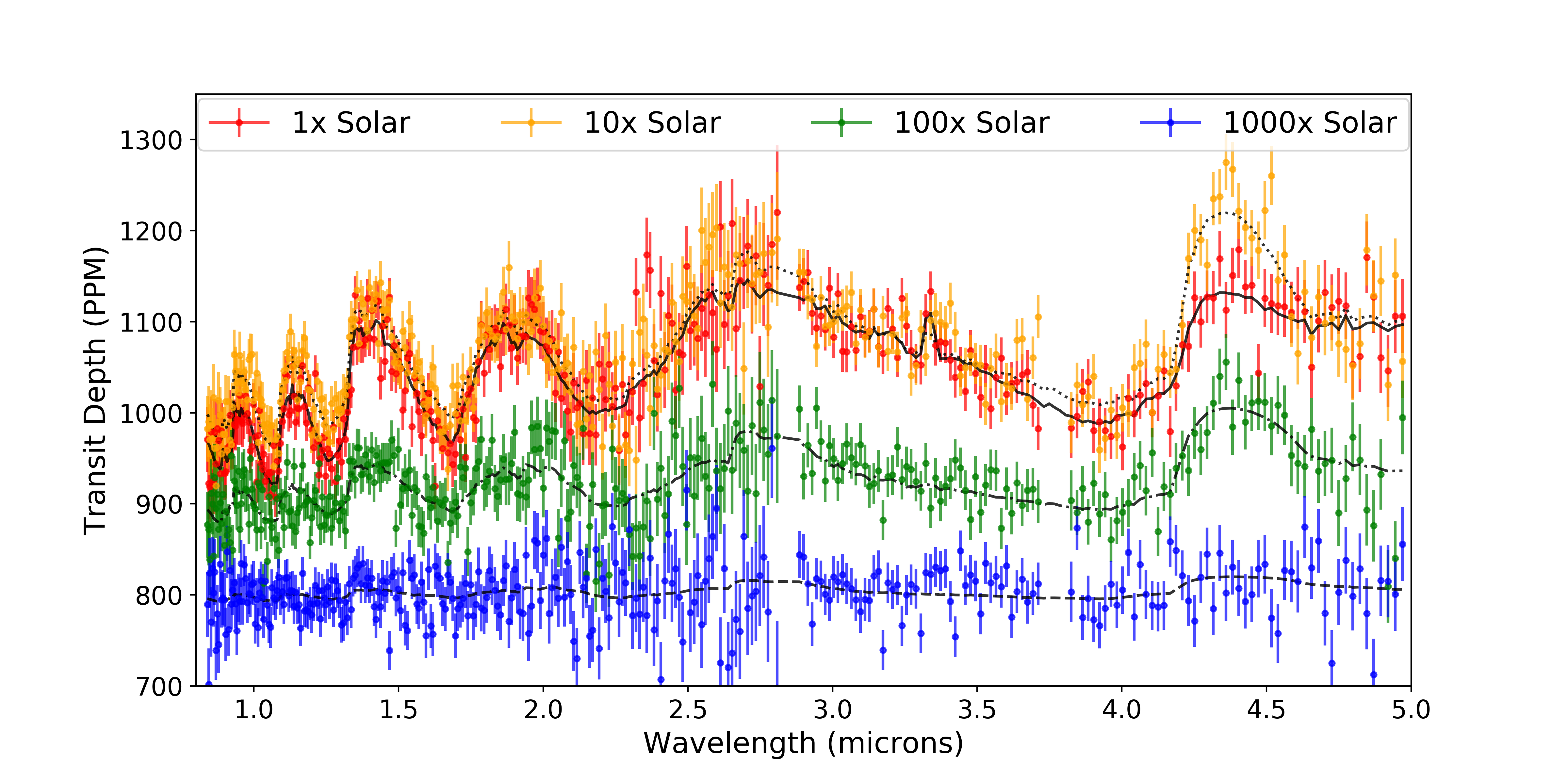}
\caption{This figure shows the simulated near-infrared transmission spectrum of HD 63935 b based on a single transit, with the NIRISS SOSS and Nirspec G395H instruments (note that this means that the full dataset shown in this plot would require two transits to obtain, one for each instrument and corresponding wavelength range). Different colors indicate different metallicities relative to the solar value. The true model spectra are shown as solid lines, while the points are simulated observations.  \label{fig:63935b-full-simulated-spectrum}}
\end{figure*}

To validate the observability of HD 63935 b, we simulated planetary transmission spectra using \texttt{CHIMERA} \citep{line2012info,line2013systematic,line2014systematic}, then simulated transit observations of the planets using \texttt{PandEXO} \citep{Batalha2017pandexo}. A sample of our simulations (cloud-free cases only) are shown in Figure \ref{fig:63935b-full-simulated-spectrum}.

\begin{figure}[h]
\includegraphics[width=0.45\textwidth]{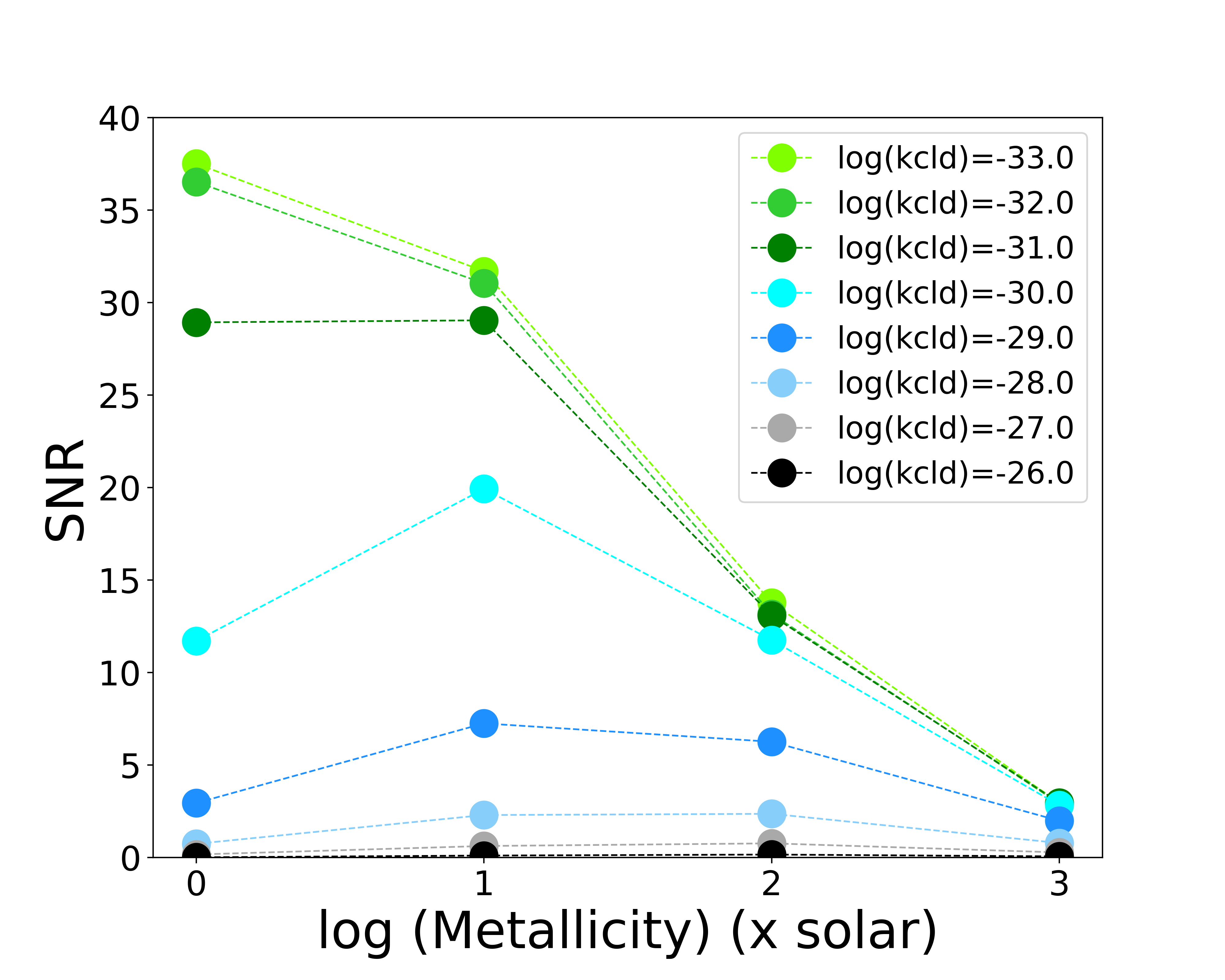}
\caption{The SNR of the 1.5 $\mu$m water feature equivalent width of HD 63935 b, calculated after a single JWST transit observation for a variety of metallicites and cloud opacities (log(kcld)). log(kcld) is a model parameter that encapsulates cloud opacity in a single greying parameter; the range used here spans the gamut from cloud-free in light green to opaque atmospheres in black. Optimistic-but-reasonable atmospheric clearness assumptions are between -30 and -29 (so SNR $\sim$10 for the 100x solar metallicity case). Obviously, an opaque atmosphere would produce no visible water features, though we describe in the text why there are reasons not to expect such a scenario. Note as well that this is a very conservative estimate of the quality of atmospheric observations. There are other water features available for obtaining a water detection, the equivalent width calculation used here ignores binning as well as more advanced retrieval techniques, and the data here come from only a single \jwst transit. SNR should scale roughly with $\sqrt{N_{obs}}$ if more transits are added.  \label{fig:waterheight_planetb}}
\end{figure}

The results for planet b support our selection of the planet as one with an exceptionally high potential for atmospheric characterization.
We simulated a range of metallicities and cloud opacities, then calculated the SNR of the water feature at 1.5$\mu$m.
We did this by simulating the spectrum 1000 times, determining the equivalent width of the feature in each, and finding the standard deviation of the resulting distribution. Our results are displayed in Figure \ref{fig:waterheight_planetb}. 
We emphasize that this is a conservative estimate of the quality of atmospheric spectra, but that even with middlingly-optimistic assumptions about metallicity and cloudiness, an observation of a single transit with \jwst can produce useful spectra of this planet, and more observations would of course produce correspondingly more precise spectra.

Additionally, there is reason to expect that the atmosphere of HD 63935 b will have high-amplitude atmospheric features in its transmission spectrum. Although cloud formation in exoplanetary atmospheres is not fully understood, extensive modelling and analysis of existing transmission spectra have been done to attempt to understand which atmospheres will be dominated by clouds or hazes. Most recently, \cite{Gao_2020} used an aerosol microphysics model to predict the dominant opacity sources in giant exoplanet transmission spectra. Their results suggest that the height of spectral features (specifically the 1.4 $\mu m$ water band) increases with increasing temperature until $\sim$950 K, with opacity in this temperature regime dominated by high-altitude photochemical hazes derived from hydrocarbons. Above $\sim$950K, silicate condensate clouds become the dominant opacity source. These clouds rise higher into the atmosphere as temperature increases, resulting in reduced spectral feature amplitude with increasing temperature from 950 until $\sim$1800K. These predictions are highly relevant to this work, as HD 63935 b sits quite close to the critical equilibrium temperature of 950K which we expect to be a local maximum of feature amplitude. 

This prediction is also in line with earlier work by \cite{CrossfieldKreidberg2017}, which analyzed the published transmission spectra of exclusively Neptune-sized exoplanets and found a positive correlation between feature height and equilibrium temperature in their domain of 500-1100K. Taken together, these results suggest that HD 63935 b is in a region of parameter space that is maximally likely to produce observable spectral features. Combined with its other desirable qualities as an atmospheric target (bright host star, large TSM value), it is clear that HD 63935 b has exceptional promise as a target for atmospheric characterization. 

Finally, we note that HD 63935 is quite sun-like (M$_* ~\sim 0.933 \textrm{M}_\odot$, [Fe/H] = 0.07 $\pm$ 0.06), making characterization of its planetary system of additional interest for comparative planetology with our own solar system. We also emphasize that only one sub-Neptune planet around such a similar star has a published atmospheric characterization study \citep[HD 3167 c][]{mikalevans2020}. Combined with other such planets with G-type host stars being discovered by TESS, it could form part of a robust population study. 

\section{Conclusions}

We have described the discovery of two planets around the bright G star HD 63935. 
\begin{itemize}
    \item HD 63935 b and c have periods $9.0600^{+0.00064}_{-0.00070}$ and $21.40 \pm 0.0019$ days, radii $2.99 \pm 0.14$ and $\crad \pm \craderr$ $R_\oplus$, and masses $\bmass \pm \bmasserr$ and $\cmass \pm \cmasserr$ \mearth. 
    \item Both confirmed planets have a radius higher than the median sub-Neptune and fall on the "radius cliff". 
    \item Planet b is an outstanding target for transmission spectroscopy, being the best target in its parameter space niche and second-best among targets on the radius cliff, while planet c is also amenable to atmospheric characterization. This quality for followup is exceptionally enticing given the compositional degeneracies that exist in this region of parameter space, which could be broken with high-quality atmospheric characterization.
    \item Atmospheric characterization of this system could provide valuable input to theories of planetary interiors, formation, and evolution, especially given that two planets similar in all observed properties except insolation flux is as close to a variable-controlled experimental setup as exoplanet astronomy typically comes. 
\end{itemize}

HD 63935 b and c attest to the bright future of exoplanet astronomy and we expect this system to be an excellent test case for studying exoplanetary atmospheres in coming years.

\textit{Facilities}
Automated Planet Finder (Levy), HARPS-N (TNG), HIRES (Keck I),  Las Cumbres Observatory Global Telescope (LCOGT), NIRC2 (Keck II), PHARO (Palomar), TESS

\textit{Software}
\texttt{AstroImageJ} \citep{Collins:2017},
\texttt{Astropy} \citep{astropy2013},
\texttt{batman} \citep{Kreidberg2015},
\texttt{emcee} \citep{emcee},
\texttt{exoplanet} and its dependencies \citep{exoplanet:exoplanet, exoplanet:agol20, exoplanet:arviz,  exoplanet:luger18, exoplanet:pymc3, exoplanet:theano},
\texttt{isoclassify} \citep{isoclassify-huber-2017},
\texttt{juliet} \citep{Espinoza2019juliet},
\texttt{Jupyter} \citep{jupyter2016},
\texttt{kiauhoku} \citep{Claytor_2020},
\texttt{matplotlib} \citep{Hunter2007matplotlib},
\texttt{numpy} \citep{numpy},
\texttt{pandas} \citep{pandas},
\texttt{Radvel} \citep{Fulton2018},
\texttt{smint} \citep{Piaulet_2021},
\texttt{SpecMatch-Syn} \citep{Petigura2017}
\texttt{Transit Least Squares} \citep{TLS}

\section{Acknowledgments}
We thank the anonymous referee for their helpful feedback that improved the quality of this work. We thank the time assignment committees of the University of California, the California Institute of Technology, NASA, and the University of Hawaii for supporting the TESS-Keck Survey with observing time at Keck Observatory and on the Automated Planet Finder. We thank NASA for funding associated with our Key Strategic Mission Support project. We gratefully acknowledge the efforts and dedication of the Keck Observatory staff for support of HIRES and remote observing. We recognize and acknowledge the cultural role and reverence that the summit of Maunakea has within the indigenous Hawaiian community. We are deeply grateful to have the opportunity to conduct observations from this mountain. We thank Ken and Gloria Levy, who supported the construction of the Levy Spectrometer on the Automated Planet Finder. We thank the University of California and Google for supporting Lick Observatory and the UCO staff for their dedicated work scheduling and operating the telescopes of Lick Observatory. 

This paper is based on data collected by the TESS mission. Funding for the TESS mission is provided by NASA's Science Mission Directorate. We acknowledge the use of public TESS data from pipelines at the TESS Science Office and at the TESS Science Processing Operations Center. This research has made use of the Exoplanet Follow-up Observation Program website, which is operated by the California Institute of Technology, under contract with the National Aeronautics and Space Administration under the Exoplanet Exploration Program. Resources supporting this work were provided by the NASA High-End Computing (HEC) Program through the NASA Advanced Supercomputing (NAS) Division at Ames Research Center for the production of the SPOC data products. This paper includes data collected by the TESS mission that are publicly available from the Mikulski Archive for Space Telescopes (MAST). 
This research has made use of the NASA Exoplanet Archive and Exoplanet Follow-up Observation Program website, which are operated by the California Institute of Technology, under contract with the National Aeronautics and Space Administration under the Exoplanet Exploration Program.
Based on observations made with the Italian Telescopio
Nazionale Galileo (TNG) operated on the island of La Palma by the
Fundaci\'on Galileo Galilei of the INAF (Istituto Nazionale di
Astrofisica) at the Spanish Observatorio del Roque de los Muchachos of
the Instituto de Astrofisica de Canarias under programmes CAT19A\_162
and CAT19A\_96.
This work has made use of data from the European Space Agency (ESA)
mission {\it Gaia} (\url{https://www.cosmos.esa.int/gaia}), processed by
the {\it Gaia} Data Processing and Analysis Consortium (DPAC,
\url{https://www.cosmos.esa.int/web/gaia/dpac/consortium}). Funding for
the DPAC has been provided by national institutions, in particular the
institutions participating in the {\it Gaia} Multilateral Agreement.
This work is partly supported by JSPS KAKENHI Grant Numbers JP17H04574 and JP18H05439, JST PRESTO Grant Number JPMJPR1775, and Grant-in-Aid for JSPS Fellows, Grant Number JP20J21872.
This work makes use of observations from the LCOGT network.

D. D. acknowledges support from the TESS Guest Investigator Program grant 80NSSC19K1727 and NASA Exoplanet Research Program grant 18-2XRP18\_2-0136.
J.M.A.M. is supported by the National Science Foundation Graduate Research Fellowship Program under Grant No. DGE-1842400. J.M.A.M. acknowledges the LSSTC Data Science Fellowship Program, which is funded by LSSTC, NSF Cybertraining Grant No. 1829740, the Brinson Foundation, and the Moore Foundation; his participation in the program has benefited this work.
M.R.K. is supported by the NSF Graduate Research Fellowship, grant No. DGE 1339067.
P.D. acknowledges support from a National Science Foundation Astronomy and Astrophysics Postdoctoral Fellowship under award AST-1903811.

\bibliography{main.bib}
\bibliographystyle{aasjournal}

\end{document}